\documentclass[traditabstract]{aa}
\usepackage{psfrag,graphicx}
\usepackage{txfonts}
\usepackage{natbib}
\usepackage{ulem}
\usepackage{color}

\begin{document}
\bibliographystyle{aa}

\title{Lyman alpha emission from the first galaxies: Implications of UV backgrounds and the formation of molecules}
\titlerunning{Lyman alpha emission from the first galaxies}

\author{M.~A.~Latif \inst{1}
\and
Dominik~R.~G.~Schleicher \inst{2,3}
\and
M.~Spaans \inst{1}
\and
S.~Zaroubi\inst{1,4} } 
\institute{ Kapteyn Astronomical Institute, University of Groningen, P.O.~Box 800, 9700 AV Groningen, The Netherlands 
\and
 Leiden Observatory, Leiden University, P.O.~Box 9513, NL-2300 RA Leiden, The Netherlands 
\and
ESO Garching, Karl-Schwarzchild-Str.2, 85748 Garching bei Munchen, Germany 
\and 
Physics Department, Technion, Haifa 32000, Israel}

\authorrunning{Latif et al.}

\date{today}






\abstract{The Lyman alpha line is a robust tracer of high redshift galaxies. We present estimates of Lyman alpha emission from a protogalactic halo illuminated by UV background radiation fields with various intensities. For this purpose, we performed cosmological hydrodynamics simulations with the  adaptive mesh refinement code FLASH, including a detailed network for primordial chemistry, comprising the formation of primordial molecules, a multi-level model for the hydrogen atom as well as the photo-ionization and photo-dissociation processes in a UV background. We find that the presence of a background radiation field $\rm J_{21}$ excites the emission of Lyman alpha photons, increasing the Lyman $\alpha$ luminosity up to two orders of magnitude. For a halo of $\rm \sim 10^{10}~M_{\odot}$, we find that a maximum flux of $\rm 5 \times 10^{-15}~erg~cm^{-2}~s^{-1}$ is obtained for $\rm J_{21} \times f_{esc}=0.1$, where $\rm f_{esc}$ is the escape fraction of the ionizing radiation. Depending on the environmental conditions, the flux may vary by three orders of magnitude. For $\rm J_{21} \times f_{esc} > 0.1$ the Lyman alpha luminosity decreases as the atomic hydrogen abundance becomes rather small. The fluxes derived here can be probed using Subaru and the upcoming James Webb Space Telescope. The emission of Lyman alpha photons is extended and comes from the envelope of the halo rather than its core. In the center of the halo, line trapping becomes effective above columns of $\rm 10^{22}~cm^{-2}$ and suppresses the emission of Lyman alpha. In addition, cooling by primordial molecules may decrease the gas temperature in the central region, which further reduces Lyman $\alpha$ emission. In the central core, $\rm H_{2}$ is photo-dissociated for a background flux of $\rm J_{21} \geq 1000$. For weaker radiation fields, i.e. $\rm J_{21}<0.1$, $\rm H_{2}$ and HD cooling are particularly strong in the center of the halo, leading to gas temperatures as low as $\rm \sim 100$ K. We also performed a parameter study with different escape fractions of ionizing photons and explored the relative role of ionizing and dissociating radiation. We find that Lyman alpha emission depends more on the strength of the ionizing background. For a constant ionizing background, the Lyman $\alpha$ flux increases at least by an order of magnitude for stronger photodissociation.}






\maketitle

\section{Introduction}

The Lyman alpha line is an important probe of the first galaxies as they are potentially strong Lyman alpha emitters. Recently, observations of galaxies have been reported at $z > 7$ \citep{2011Natur.469..504B,2010MNRAS.409..855B, 2010arXiv1011.5500V}. The upcoming James Webb Space Telescope (JWST) will further enhance our ability to detect galaxies at higher redshift by exploring them with higher sensitivity and test our current understanding of early structure formation. The detection of the Lyman alpha line will be a key feature in the observation of galaxies in the early universe.

Numerous Lyman alpha emitters have been detected so far at intermediate redshifts \citep{2010Natur.467..940L,2000ApJ...532..170S,2009ApJ...696.1164O,2008ApJ...675.1076S,2006ApJ...648...54S,2009ApJ...693.1579Y,2004AJ....128..569M}.  The origin and source of these blobs is still not completely known. The energy sources of these Lyman alpha blobs include photo-ionization by stars or miniquasars \citep{2001ApJ...556...87H,2006Natur.441..120J}. Another potential driver of these emitters could be the emission from cold streams and accretion flows \citep{2006A&A...452L..23N,2007MNRAS.378L..49S}. The accreted gas releases its binding energy in the gravitational potential of the halo, which results in the generation of shocks and the emission of Lyman alpha photons. This scenario has also been supported by theoretical studies and numerical simulations \citep{2000ApJ...537L...5H,2006ApJ...649...37D,2006ApJ...640..539Y,2006ApJ...649...14D,2008ApJ...682..745W,2009MNRAS.400.1109D,2009Natur.457..451D,2010MNRAS.407..613G,2010arXiv1011.0438P,2010ApJ...725..633F,2011MNRAS.411.1659L}. Some of the observed Lyman alpha blobs have been associated with massive star forming galaxies \citep{2006ApJ...640L.123M}.


In our previous study \citep{2011MNRAS.tmpL.217L}, we estimated the emission of Lyman alpha photons from a protogalactic halo in the absence of UV radiation. We found that the emission of Lyman alpha radiation is extended and originates from accretion flows and virial shocks. The situation is however more complicated, due to the presence of ionizing and photodissociating backgrounds, as well as the formation of molecules \citep{2007ApJ...665...85J}. These are the effects we explore here.

The intense Lyman-Werner background flux can photo-dissociate $\rm H_{2}$ molecules in atomic cooling halos with virial temperature $\rm > 10^{4}$ K \citep{2001ApJ...548..509M,2001ApJ...546..635O,2007ApJ...665...85J,2008ApJ...673...14O,2008MNRAS.387.1021G,2008ApJ...686..801O,2008MNRAS.391.1961D,2010MNRAS.402.1249S}. It is found that the strength of the UV flux below $\rm J_{21}=10^{3}$ only delays the collapse and does not quench the formation of $\rm H_{2}$.  \cite{2010ApJ...712L..69S} confirmed the previous results and found that for $\rm J_{21}> 10^{3}~ H_{2}$ does not form, but the gas temperature nevertheless declines down to 5000 K due to cooling from higher electronic transitions of hydrogen atom. 

The implications of ionizing radiation for the chemistry in the halo are, on the other hand, much more unexplored. During the epoch of reionization, background UV radiation generated by stellar populations will photoionize the intergalactic gas \citep{2000ApJ...535..530G,2008PhRvD..78h3005S,2008MNRAS.384.1080T,2008ApJ...684....1W,2009MNRAS.393..171M}. The recombination following the absorption of stellar photons may further boost the Lyman alpha emission from accretion flows. \cite{2005ApJ...622....7F} found that the photoionizing background can power the Lyman alpha emissivity and can explain the observed Lyman alpha blobs. Our main goal here is to study the impact of background UV on Lyman alpha emission and to see how it influences the spatial emission profile.

Gas ionized by UV radiation cools faster than it recombines, preserving an increased ionization degree. The free electrons act as catalysts in the formation of molecular hydrogen \citep{1967Natur.216..976S,1998A&A...335..403G,2007ApJ...663..687Y}. In the absence of dust, H$_2$ formation proceeds via the gas phase, with the dominant formation channel being due to H$^-$:
\begin{equation}
\rm H + e^{-} \rightarrow H^{-} + \gamma .
\label{h2}
\end{equation}
This then leads to H$_2$ formation via the reaction
\begin{equation}
\rm H + H^{-} \rightarrow  H_{2} + e^-.
\label{h21}
\end{equation}

Another prominent coolant in primordial gas is HD. Its dominant formation path is via collisions between $\rm H_{2}$ molecules and ionized deuterium:
\begin{equation}
\rm H_{2} + D^{+} \rightarrow  HD + H^{+}.
\label{h22}
\end{equation}
While  $\rm H_{2}$ molecules can cool the gas to a few hundred Kelvin, HD can decrease the temperature even further if it is abundant enough. While the H$_2$ molecule is perfectly symmetric and thus has no permanent dipole moment, the deuterium breaks the symmetry of the molecule, thus creating a small but non-negligible dipole moment. This way, the low energy dipole transitions are no longer forbidden, and the spontaneous emission rates are generally increased.

In order to investigate the influence of stellar UV radiation on the emissivity of Lyman alpha photons, we have used state-of-the-art high resolution cosmological simulations and include for the first time the detailed chemistry of all relevant processes for primordial gas (i.e., photo-ionization, photo-dissociation, collisional ionization, collisional and radiative recombination and the formation of molecules), treating atomic hydrogen as a multi-level atom and including the effects of line trapping. In this paper, we present a number of simulations with different strengths of background radiation fields and escape fractions of ionizing photons to explore the relative importance of ionizing and photo-dissociating radiation. We follow the collapse of a zero metallicity halo and compute the Lyman $\alpha$ emissivity under different conditions.

This paper is structured as follows. In section 2, we briefly discuss the simulation setup and numerical methods used. In section 3, we give an overview of our chemical model. In section 4, we present the results obtained.  Finally, in section 5 we discuss our conclusions.

\section {Numerical Methodology}

To investigate the role of background radiation fields in the emission of Lyman alpha photons from protogalactic halos, we use the code FLASH \citep{Dubey2009512}. FLASH is an adaptive mesh, parallel simulations code. It is a module-based Eulerian grid code which can solve a broad range of astrophysics problems. It makes use of the PARAMESH library to handle block structured adaptive grids. It uses the message passing interface (MPI) library to achieve portability and scalability on many different parallel systems. The adaptive mesh technique is used to achieve high dynamic resolution with minimum computational time. We use adaptive mesh refinement (AMR) to add resolution in the regions of interest. We use an unsplit hydro solver with the 3rd order piece-wise parabolic method (PPM) for hydrodynamic calculations \citep{1984JCoPh..54..174C}. This method is well suited for flows involving shocks and contact discontinuities. A multigrid solver is used to solve for self-gravity \citep{2008ApJS..176..293R}. The dark matter is simulated based on the particle mesh (PM) method.

We run the COSMICS package developed by \cite{1995astro.ph..6070B} to produce Gaussian random field initial conditions. We make use of constraint realizations implemented in the grafic code (part of the COSMICS package) to select a massive halo. We start our simulation at redshift 100 as determined by COSMICS. Our computational box has a comoving size of 10 Mpc. We use periodic boundary conditions both for hydrodynamics and gravity. We start the simulations with an initial resolution of $\rm 512^{3}$ grid cells in the central 1 Mpc region and set the rest of the box to a resolution of $\rm 128^{3}$ grid cells. We initialize $\rm 2.6 \times 10^{6}$ particles to compute the evolution of dark matter. For the cosmology, we adopt the $\rm \Lambda$CDM model with the WMAP 5-year parameters ($\rm \Omega_{m} =0.2581$, $\rm H_{0}=72~km~s^{-1}~Mpc^{-1}$, $\rm \Omega_{b}=0.0441$) and a scale-invariant power spectrum with the preferred value of $\rm \sigma_{8}=0.8$. We select the most massive halo and follow its collapse using the AMR method. We impose 7 additional levels of refinement, which gives 15 levels of refinement in total. This corresponds to a dynamical resolution limit of 75 pc in comoving units. We resolve the Jeans length with a minimum of 20 cells. This should be enough to properly resolve the essential physics, for further details about the significance of resolution criteria see \cite{2011arXiv1102.0266F}. This automatically ensures that we also fulfill the Truelove criterion to suppress artificial fragmentation \citep{1997ApJ...489L.179T}.

\begin{table*}[htb]

\begin{center}
\caption{List of simulations presented in this paper}
\begin{tabular}{ccccc}
\hline
\hline

Nr.	& UV flux				& Escape fraction	   & Lyman alpha Luminosity			& Lyman alpha flux	\\

	 & $\rm J_{21} [erg/cm^{2}/s/Hz/sr]$	& 	                   & [$erg/s$]	         & $\rm [erg/cm^{2}/s]$	 \\ 

\hline
1	& 0.01					& 0.01		          & $10^{43}$			       & $5 \times 10^{-18}$  \\		
2	& 0.1					& 0.01		          & $10^{44}$		               & $10^{-17}$		\\
3	& 10					& 0.01		          & $10^{46}$		               & $5 \times 10^{-15}$	 \\	
4	& 100					& 0.01		          & $5 \times10^{45}$		       & $10^{-15}$		\\
5       & 1000                                  & 0.01                    & $10^{45}$                             & $2 \times 10^{-16}$  \\
6       & 0.01                                  & 0.1                     & $10^{44}$                             & $4 \times 10^{-17}$  \\
7	& 0.1                                   & 0.1                     & $10^{46}$                             & $5 \times 10^{-15}$   \\
8       & 0.01                                  & 0.5                     & $10^{46}$                             & $5 \times 10^{-15}$   \\
9       & 0.1                                   & 0.5                     & $10^{46}$                             & $5 \times 10^{-15}$    \\

\hline
\end{tabular}
\label{tab:sumary}
\end{center}

\end{table*}

\section{Chemistry and Cooling}

Primordial gas is mainly composed of hydrogen and helium. Microphysical processes of these atoms play a vital role in the formation of the first structures. Stars and galaxies are formed in protogalactic halos where collapse is induced due to cooling by hydrogen lines. Lyman alpha line cooling is the dominant process in metal free gas for temperatures down to 8000 K.  $\rm H_{2}$ and HD are the only relevant coolants at temperatures below $\rm 10^{4}$ K \citep{1998A&A...335..403G, 1984ApJ...280..465L}. \cite{1967Natur.216..976S} were first to realize the importance of molecular hydrogen formation through gas phase reactions. They found that trace amounts of $\rm H_{2}$ can be formed through gas-phase reactions already at $z\sim300$. The rotational and vibrational modes of molecular hydrogen can be excited at low temperatures and cool the gas to a few hundred Kelvin. HD is an important coolant around 100 K and can cool the gas to lower temperatures.

We expect that the metallicity in the IGM may still be close to primordial at high redshift, in particular for isolated halos. For Lyman $\alpha$ emission driven by accretion and infall, adopting a primordial composition thus allows us to capture the dominant effects regulating the emission process. Even for metallicities up to 1\% solar, cooling procedes mostly via primordial coolants for the densities in our simulations \citep{2005ApJ...626..627O,2007ApJ...660.1332J}. Depending on the properties of dust grains, they could potentially enhance H$_2$ formation \citep{2009A&A...496..365C}. However, we assume here that either the grain-size distribution is such that this effect is not significant, or that the additional H$_2$ formation is suppressed by the photodissociating background. The effects of dust will be explored in more detail in an upcoming paper.

To study the thermal evolution of the gas, it is crucial to model the chemistry of primordial gas in detail. Solving the rate equations self-consistently along with hydrodynamics is a  computationally demanding task due to the stiff nature of the reaction network. We devised a chemical network consisting of 36 reactions which involve 23 collisional and 13 radiative processes. The reactions and rate coefficients for these processes and their references  are listed in table 1 of appendix A. We solve the rate equations of the following 12 species  $\rm H,~H^{+},~He,~He^{+},~He^{++},~e^{-},~H^{-},~H_{2},~H_{2}^{+},~D,~D^{+},~HD$. Our network is based on \cite{1997NewA....2..209A,1997NewA....2..181A,2008A&A...490..521S,2007ApJ...666....1G}, see Appendix A, and includes collisional ionization, radiative recombination, photo-ionization and photo-dissociation rates of all the species. The rate equations for these species are solved using the backward differencing scheme (BDF), see \cite{1997NewA....2..209A}. The BDF method is not fully implicit but still more accurate and stable than other schemes. We have implemented a comprehensive model for cooling and heating of primordial gas. It comprises collisional excitation cooling, collisional ionization cooling, recombination cooling, Bremsstrahlung cooling, Compton cooling/heating, photo-ionization heating and photo-dissociation heating, as well as a multi-level model for atomic hydrogen line cooling. We also included the cooling due  to $\rm H_{2}$ and HD molecules. The heating / cooling rates are listed in Table 2 of appendix A. Modeling details of the chemical network are summarized in appendix A.

Another important ingredient of our modeling is the effect of line trapping of Lyman alpha photons. Lyman alpha emission is an efficient cooling process for halos with virial temperatures $\rm > 10^{4}$ K. It is oftenly presumed that gas remains effectively optically thin to Lyman alpha cooling, but large hydrogen column densities produce large optical depths to these lines. At optical depths $\rm \tau_{0} > 10^{7}$, the photon escape time typically becomes longer than gas free fall time. Because of the weak dependence of photon escape time on the gas number density, $\rm T_{ph} \propto n^{-1/9}$, as compared to free fall time, $\rm T_{ff}=1.0/(G\rho)^{1/2}$, trapping becomes effective during the process of collapse \citep{2006ApJ...652..902S}. A detailed description of
our treatment for Lyman alpha trapping and the modeling of the hydrogen line transitions is given in Appendix A.

\cite{2010ApJ...712L..69S} found that in the presence of Ly $\alpha$ line trapping, total cooling is compensated as it proceeds through the other transitions of atomic hydrogen. The transition from 2p-1s produces Lyman alpha photons while the transition from 2s-1s produces two continuum photons. These continuum photons are not trapped and cooling proceeds through them, depending on the ambient density. To model all these mechanisms, we have treated atomic hydrogen as multi-level atom by including up to the fifth excited state. We have also considered the electronic transitions from higher levels of hydrogen.

\begin{figure*}[htb!]
\centering
\begin{tabular}{c c}
\begin{minipage}{8cm}
\includegraphics[scale=0.35]{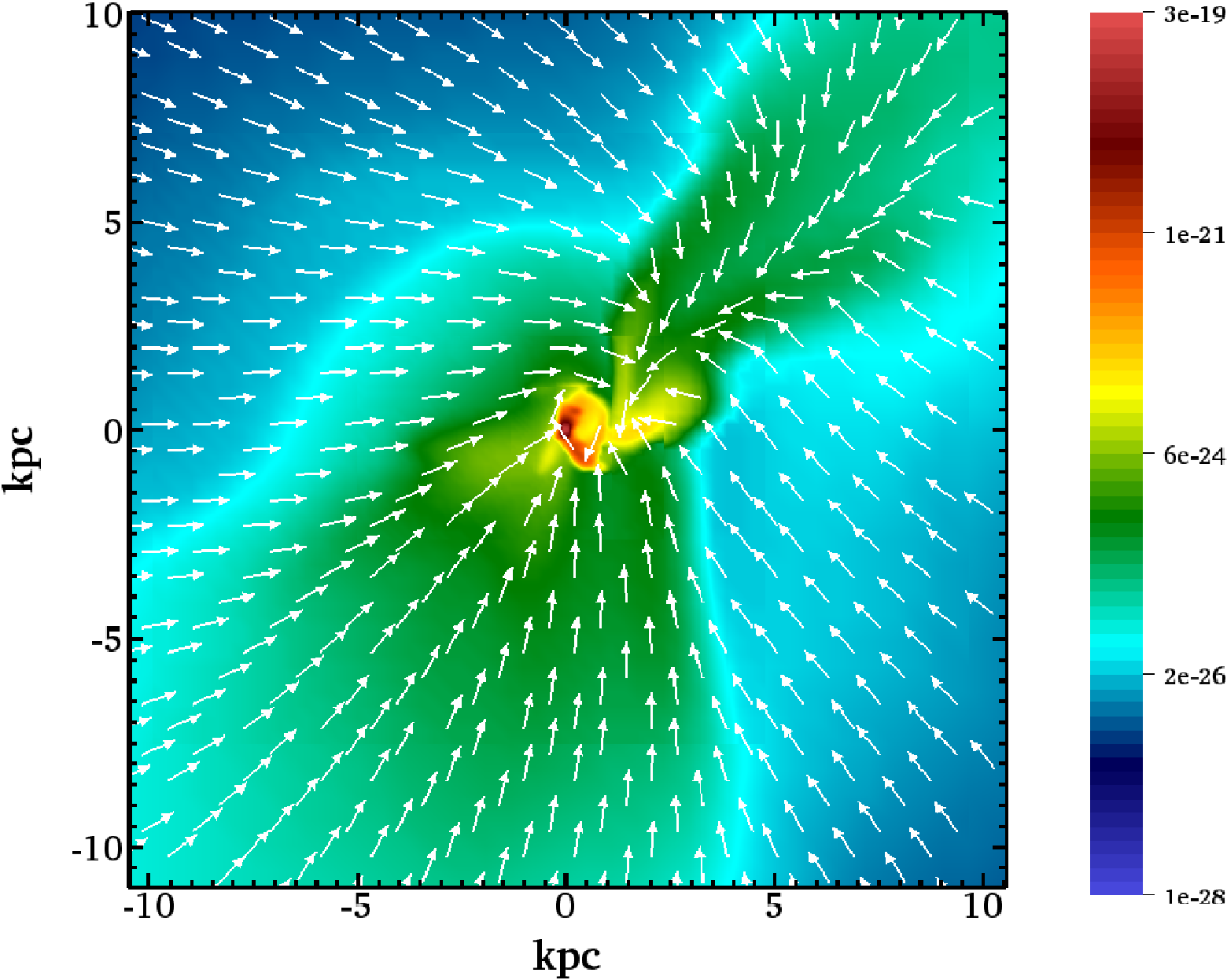}
\end{minipage} &
\begin{minipage}{8cm}
\includegraphics[scale=0.35]{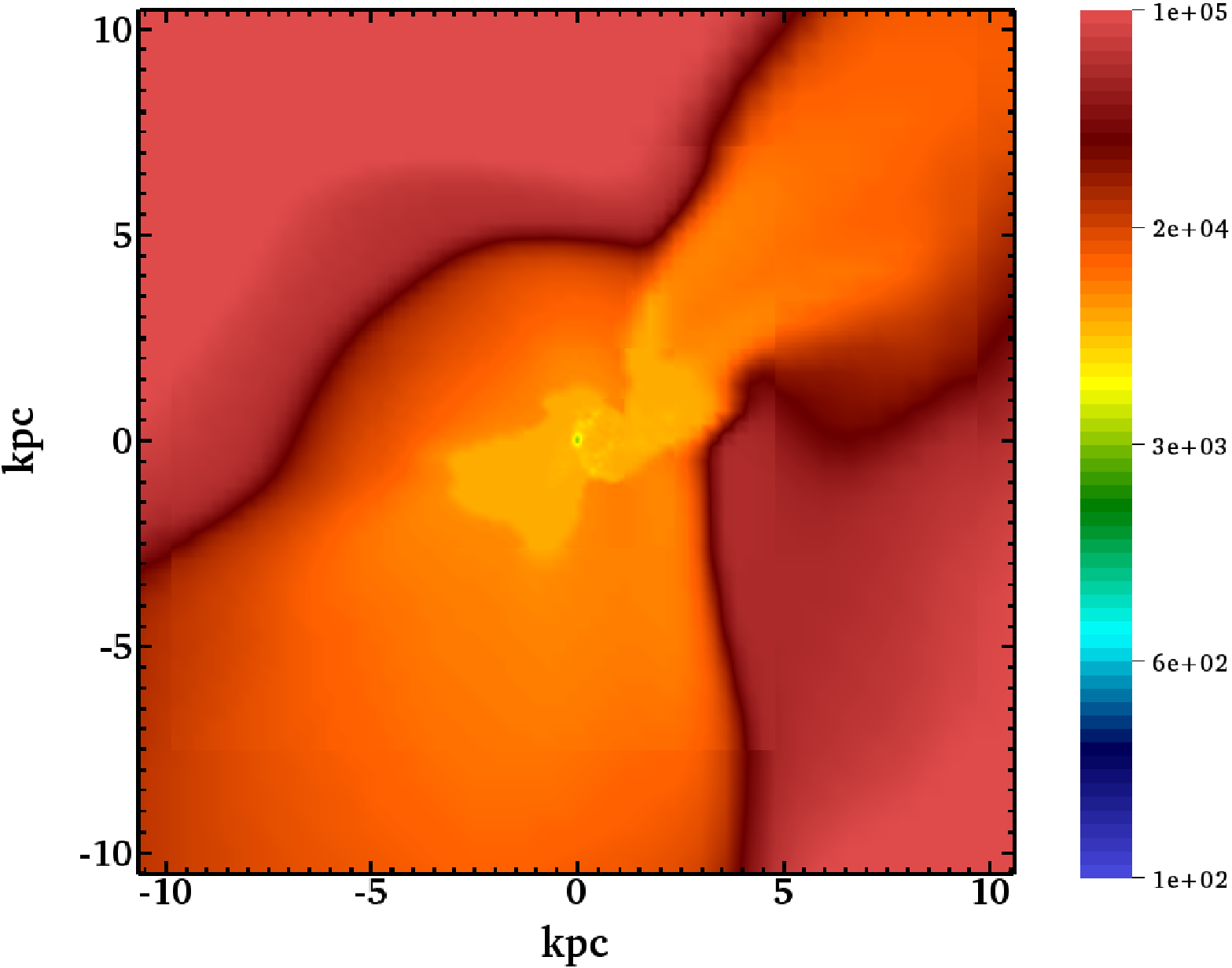}
\end{minipage} \\  \\

\begin{minipage}{8cm}
\includegraphics[scale=0.35]{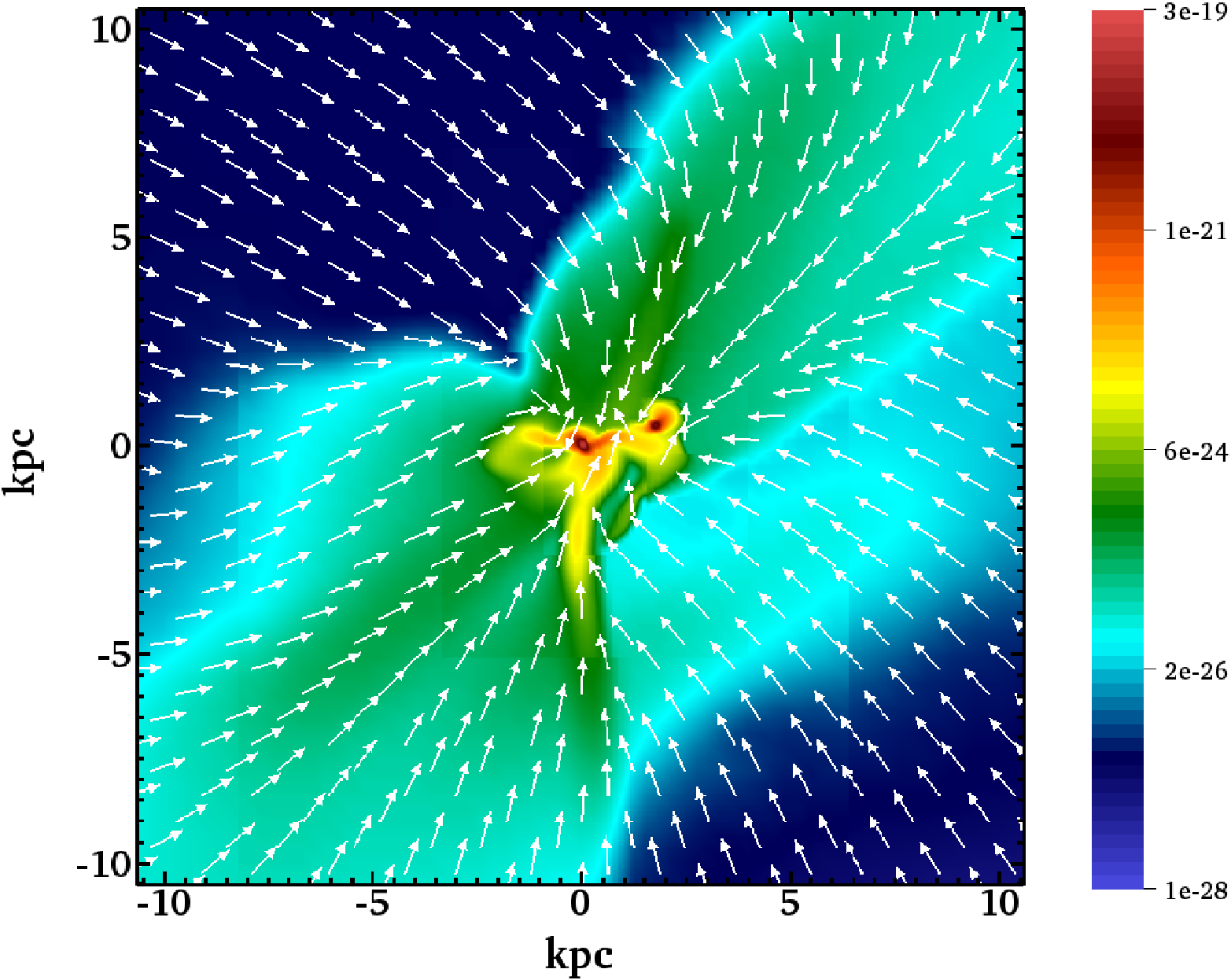}
\end{minipage} &

\begin{minipage}{8cm}
\includegraphics[scale=0.35]{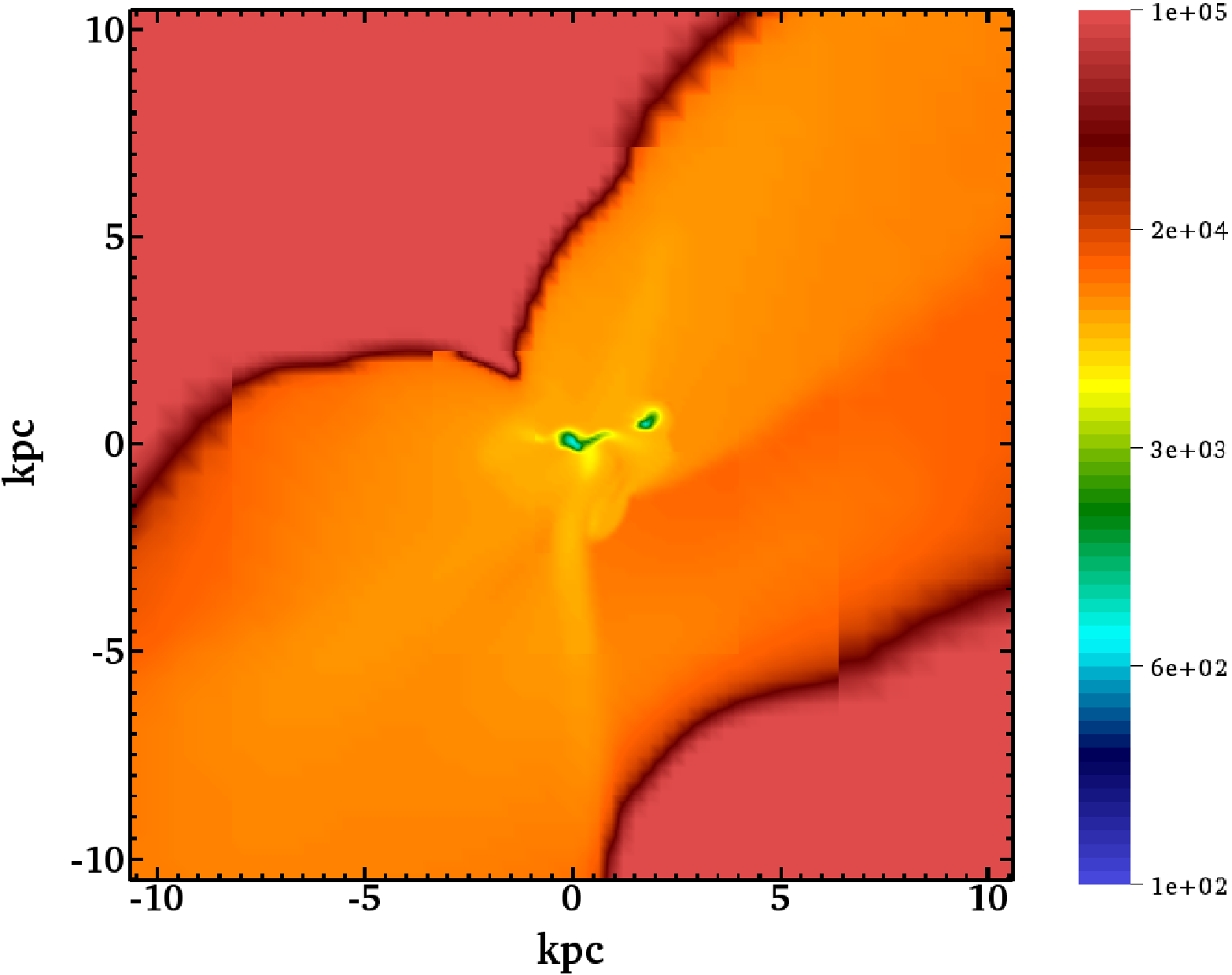}
\end{minipage} \\ \\

\begin{minipage}{8cm}
\includegraphics[scale=0.35]{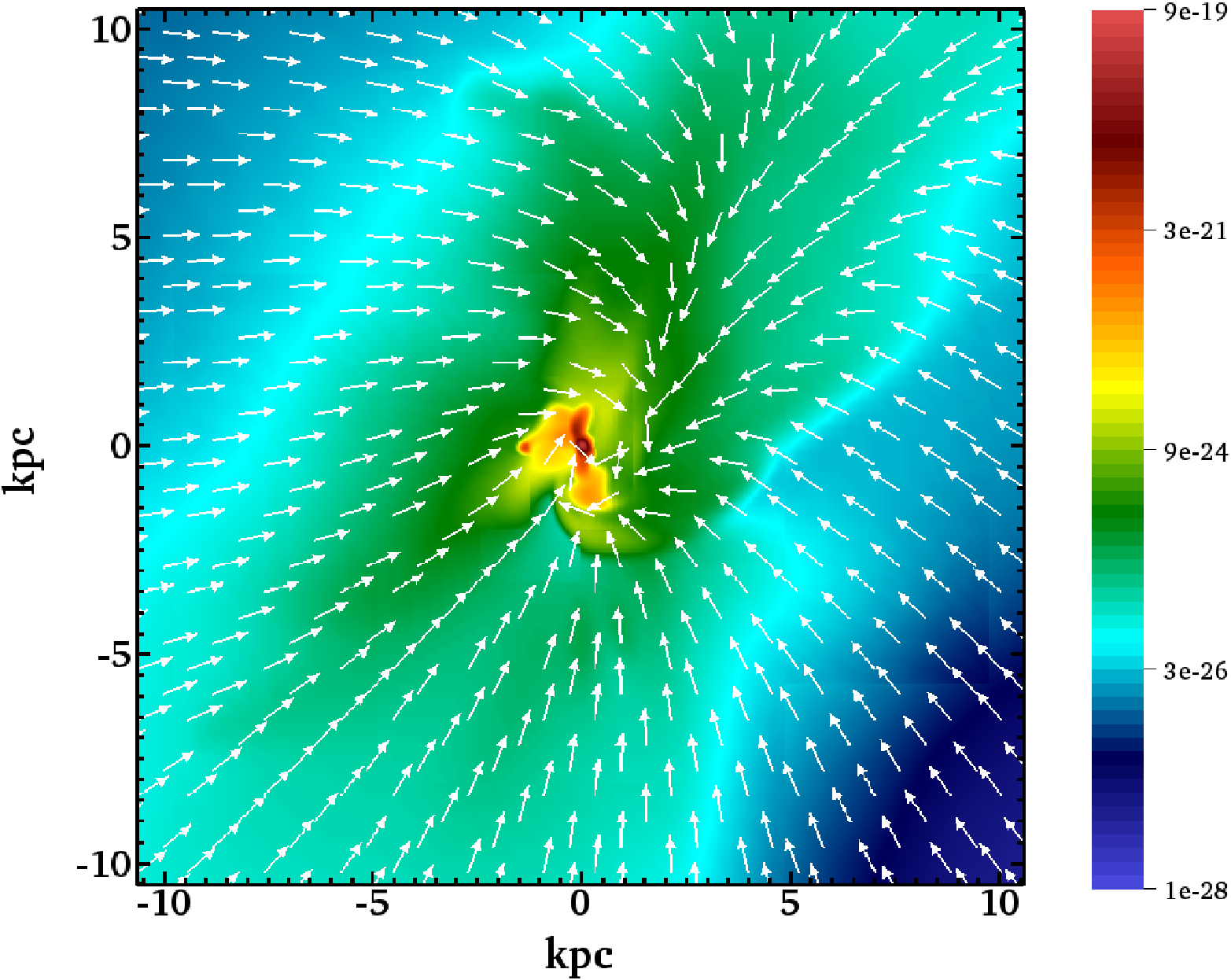}
\end{minipage} &

\begin{minipage}{8cm}
\includegraphics[scale=0.35]{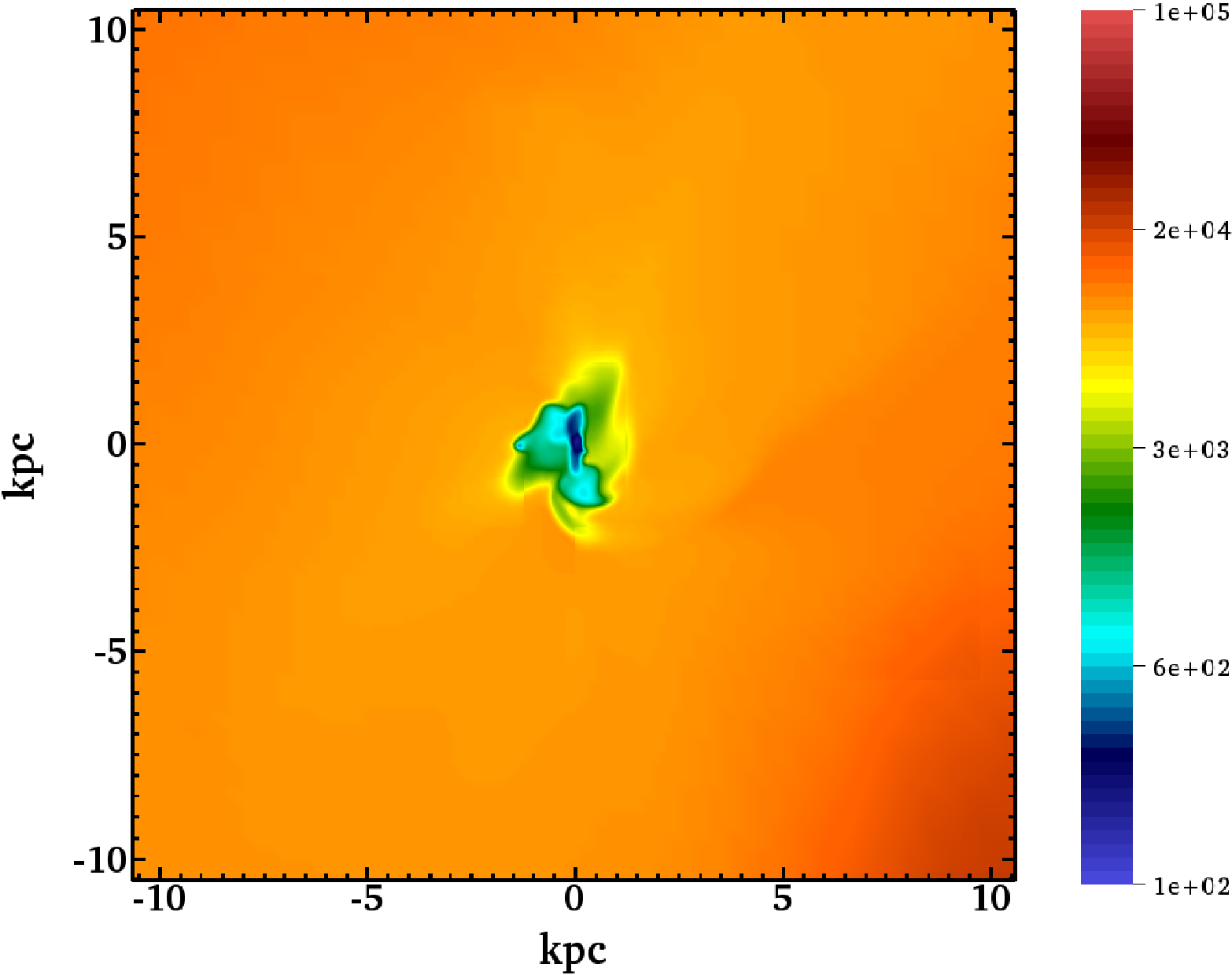}
\end{minipage} \\ \\

\end{tabular}
\caption{ The panels in this figure show the density and temperature slices at redshift 5.5. The left-hand panels show density slices through the center of the halo for $\rm J_{21}=1000,~10,~0.1$ from top to bottom and the right-hand panels show the corresponding temperature slices. Velocity vectors are overplotted on the density slices. The figure shows the inner region of 20 kpc in comoving units. All plots shown here are for an escape fraction of 1\%.}
\label{figure3}
\end{figure*}

\begin{figure*}[htb!]
\centering
\begin{tabular}{c c}
\begin{minipage}{8cm}
\includegraphics[scale=0.28]{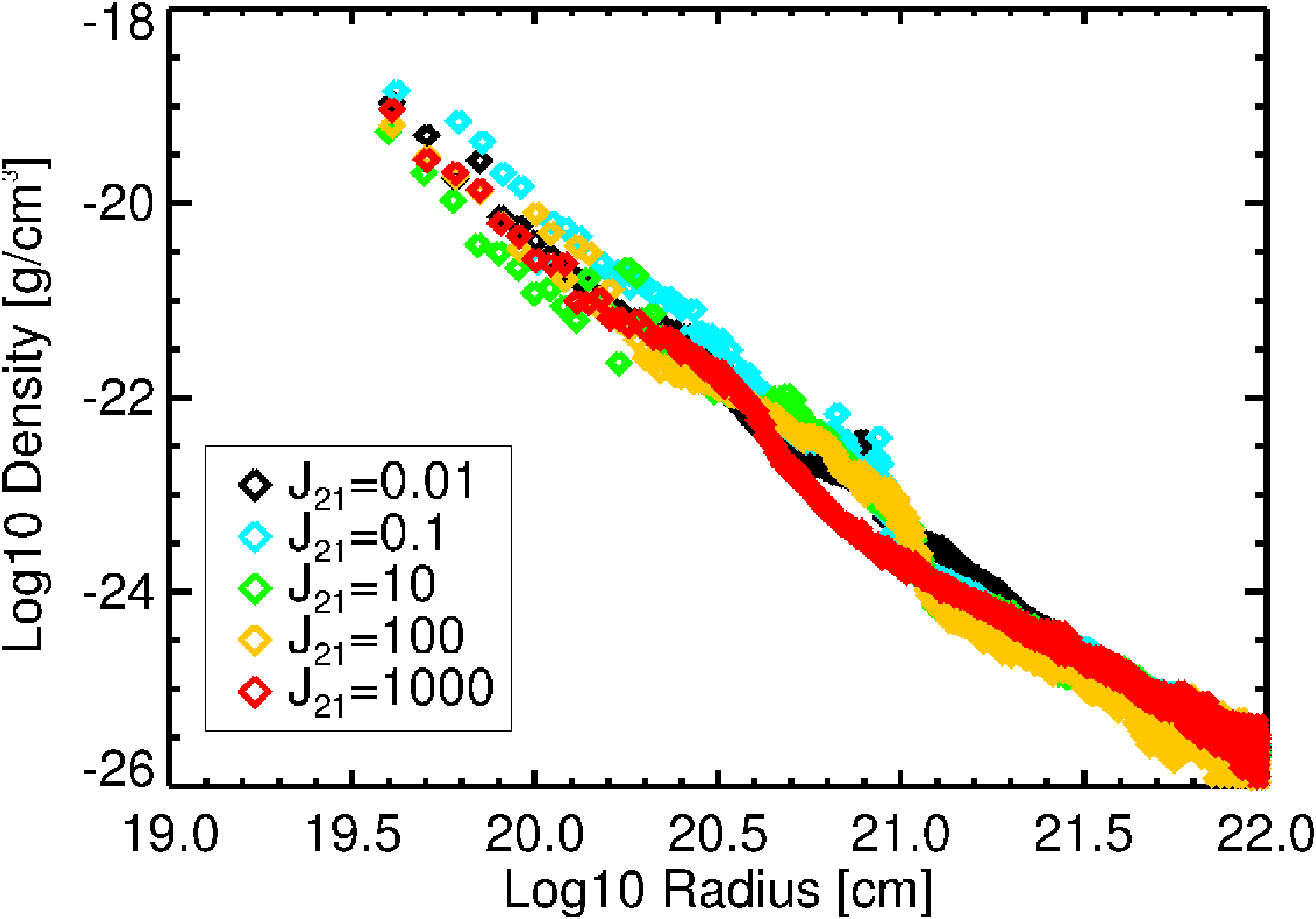}
\end{minipage} &
\begin{minipage}{8cm}
\includegraphics[scale=0.28]{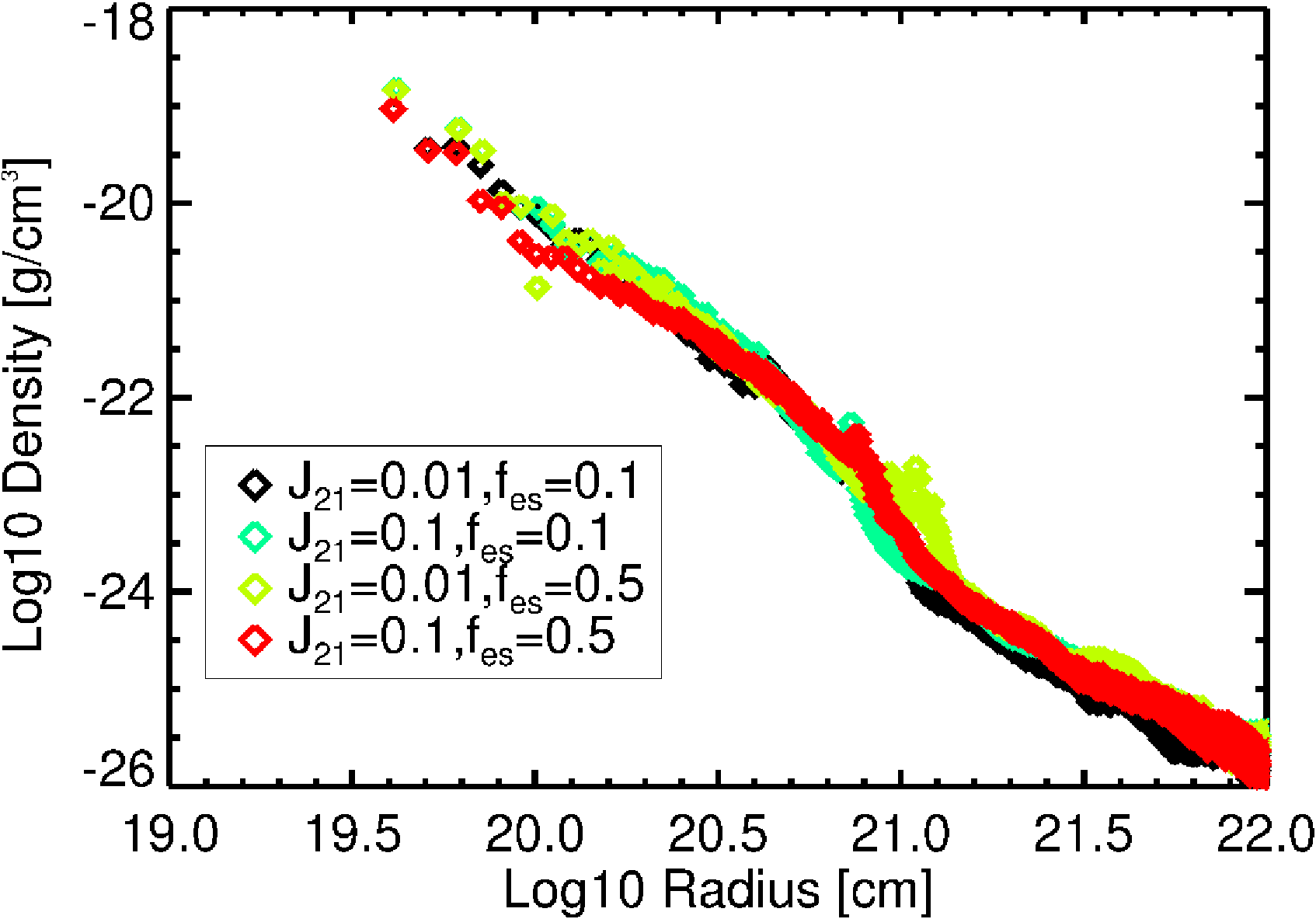}
\end{minipage} 
\end{tabular}
\caption{Figure shows the density radial profiles for different escape fractions. The left panel shows density profiles for an escape fraction of 1\%. The right panel shows the density radial profile for escape fractions of 10\% and 50\%.}
\label{figure4}
\end{figure*}

The radiation produced during the reionization process can have important implications for subsequent structure formation. These photons cannot only photo-heat and ionize the gas but also can photo-dissociate the $\rm H_{2}$ and HD molecules. The input spectrum is a blackbody with $\rm T_{*}=10^{4}$K, with the intensity above 13.6 eV reduced by the escape fraction $\rm f_{esc}$ as \citep{2008ApJ...686..801O}
\begin{equation}
\rm J_{\nu}^{ex} = J_{21}10^{-21}[B_{\nu}(T_{*})/B_{\nu H}] f(\nu)~ erg~cm^{-2}sr^{-1}s^{-1}Hz^{-1}
\label{jrad}
\end{equation}

where $\rm J_{21}$ is the mean intensity of radiation at the Lyman limit, $\rm T_{*}$ is the color temperature of a star. $\rm f(\nu)$ is one for non-ionizing photons and equal to $\rm f_{esc}$ for ionizing photons. $\rm B_{\nu}(T)$ is the stellar radiation intensity (i.e. black body spectrum),
\begin{equation}
\rm B_{\nu}(T) = {2h \nu ^{3} \over c^{2}} {1 \over (exp(h\nu/kT) -1) }.
\label{bnew}
\end{equation}

where h is Planck constant, $\rm \nu $ is the frequency of radiation, k is Boltzmann constant and T is the radiation temperature. $\rm J_{IH}=J_{21} \times f_{esc}$ represents the ionizing soft UV flux and $\rm J_{21}$ is the background UV flux below the Lyman limit (i.e. 13.6 eV) in units of $\rm 10^{-21}~erg/cm^{2}/s/Hz/sr$. While propagating through the intergalactic medium, the photons are attenuated as the optical depth depends on the frequency. In order to approximately account for the effect of self-shielding of gas in the halo, we estimate the optical depth as a function of  frequency as 
\begin{equation}
\rm \tau_{i}(\nu _{j}) = \Sigma_{i} \sigma_{i}(\nu_{j}) \times n_{i} \times \lambda_{J}
\label{taaau}
\end{equation}
where i denotes the species, j the frequency bin, $\rm \sigma$ the photo-ionization cross-section of the given species, $\rm n_{i}$ its number density and $\rm \lambda_{J}$ is the local Jeans length. For a given optical depth, attenuation of the radiation field is calculated as

\begin{equation}
\rm J^{att} (\nu) = J^{ext}(\nu) \times exp(-\tau(\nu_{i})) .
\label{bnew2}
\end{equation}

\begin{figure*}[htb!]
\centering
\begin{tabular}{c c}
\begin{minipage}{8cm}
\hspace{-0.6cm}
\includegraphics[scale=0.29]{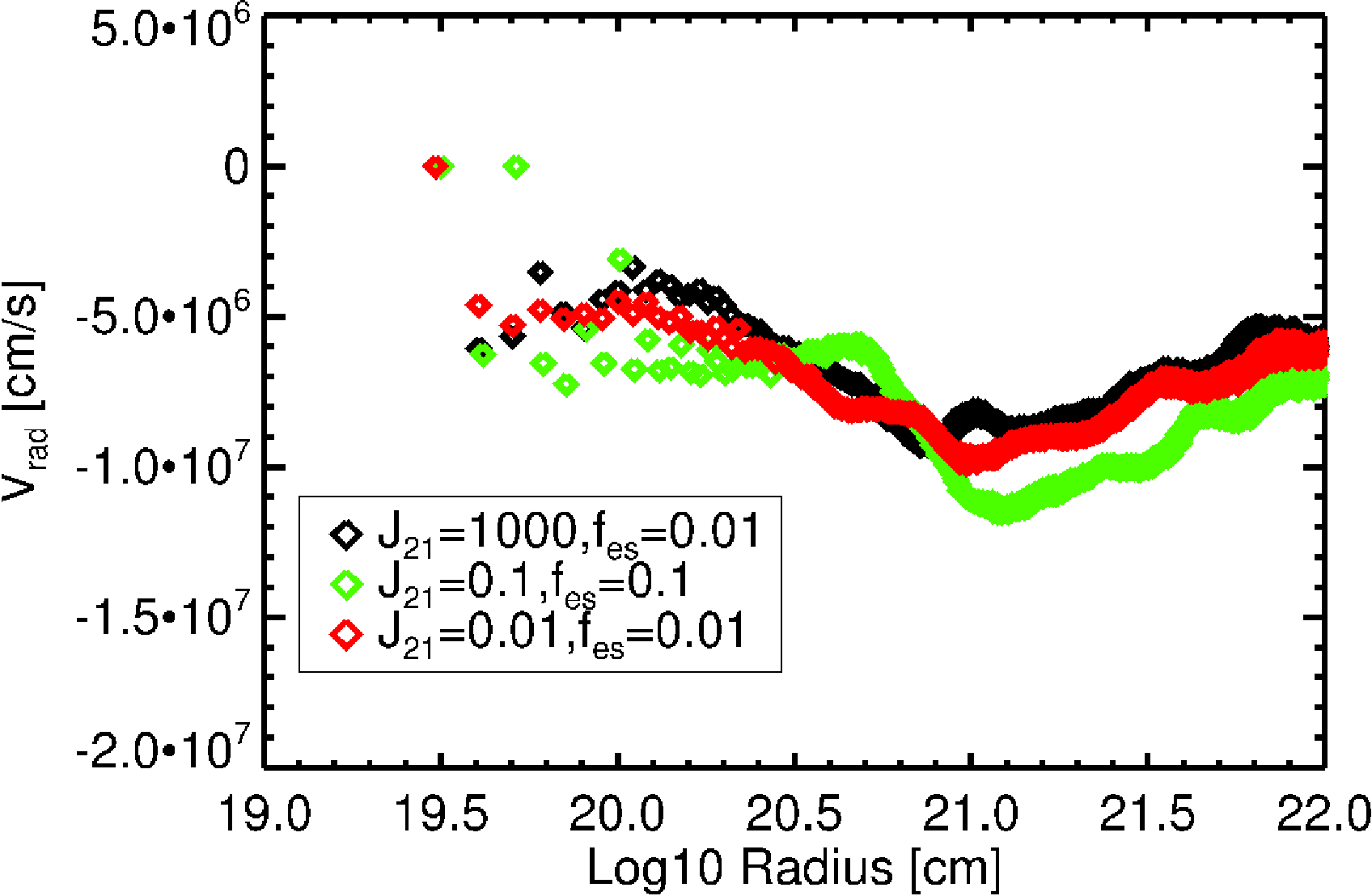}
\end{minipage} &
\begin{minipage}{8cm}
\includegraphics[scale=0.29]{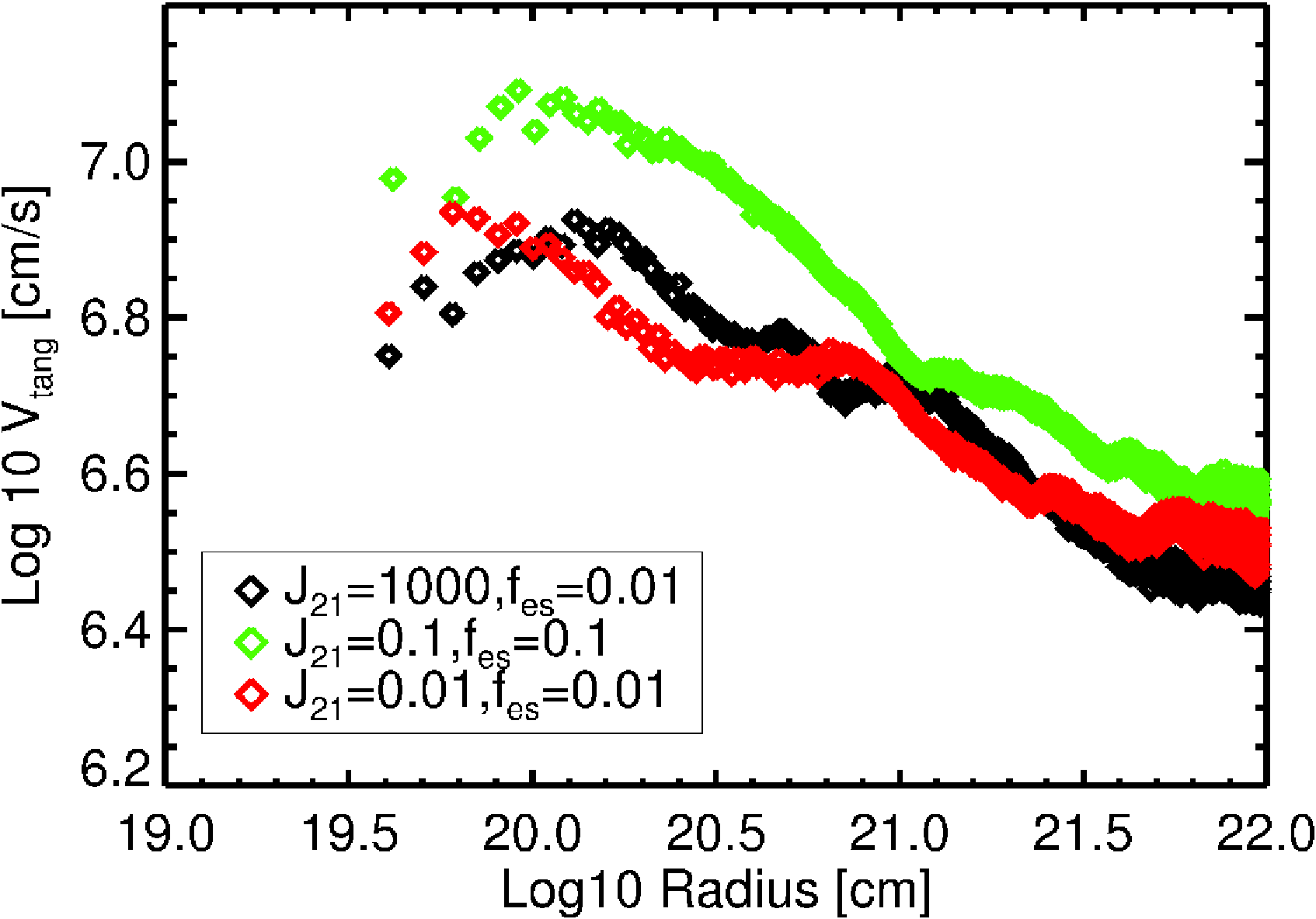}
\end{minipage} 
\end{tabular}
\caption{Left panel shows the radial velocity profile for different background UV radiation fields as given in the legend. Right panel shows the tangential velocity radial profile of the halo for different background UV fields. For the tangential velocity, the average was calculated over the absolute values. So, it is a measure of turbulence rather than rotation.}
\label{figvel}
\end{figure*}

\begin{figure*}[htb!]
\centering
\begin{tabular}{c c}
\begin{minipage}{8cm}
\hspace{0.27cm}
\includegraphics[scale=0.28]{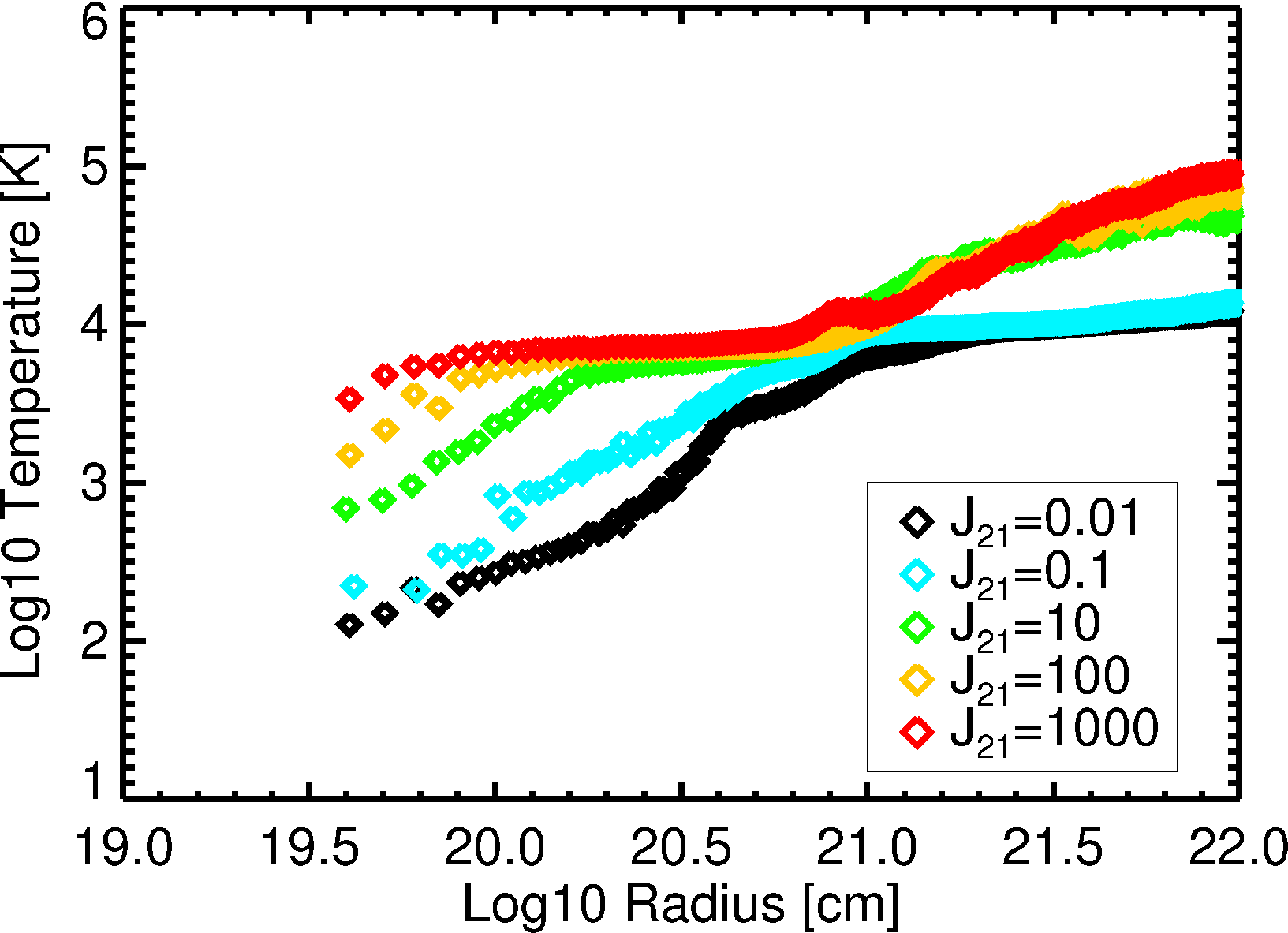}
\end{minipage} &
\begin{minipage}{8cm}
\includegraphics[scale=0.28]{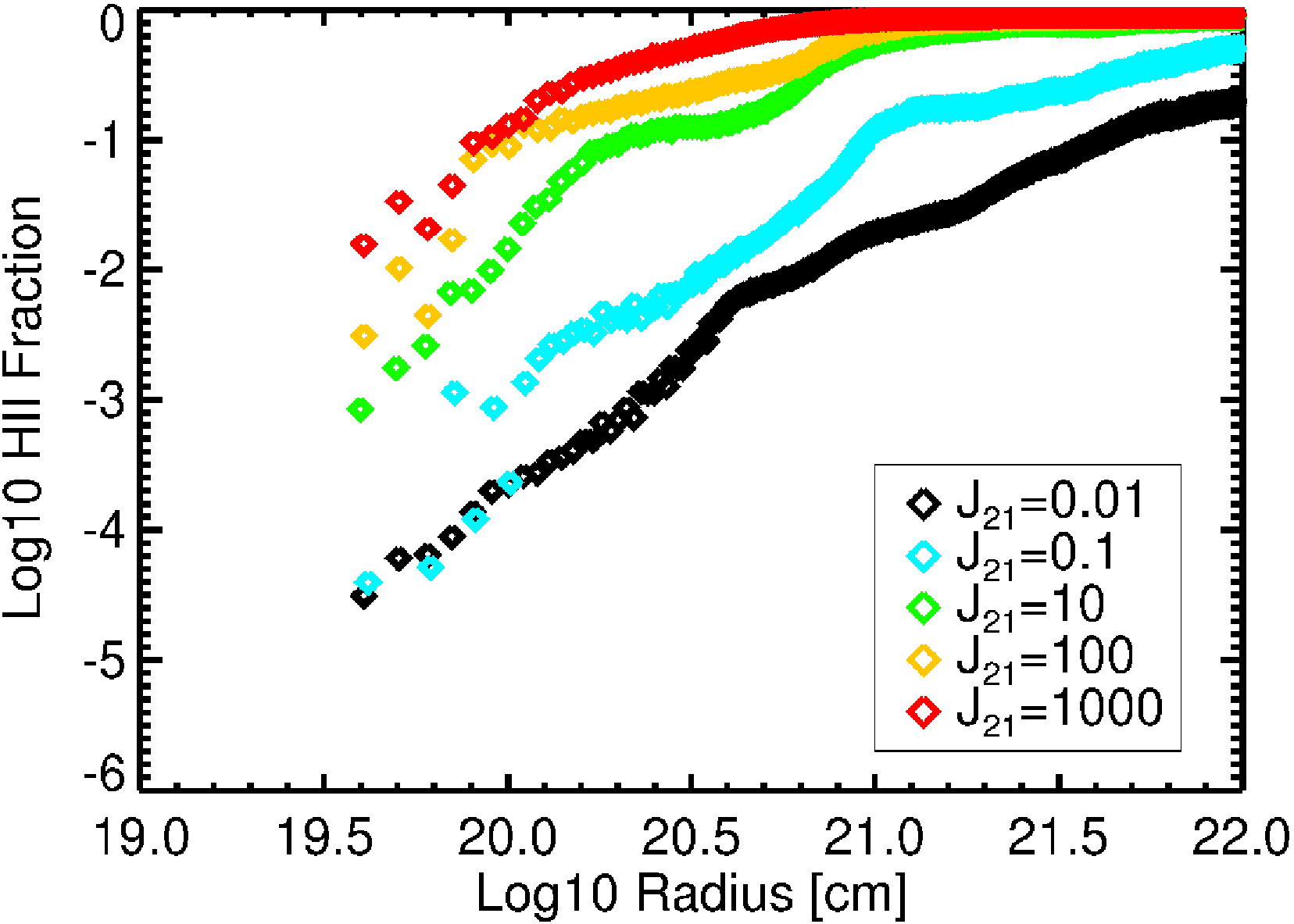}
\end{minipage} \\  \\

\begin{minipage}{8cm}
\includegraphics[scale=0.28]{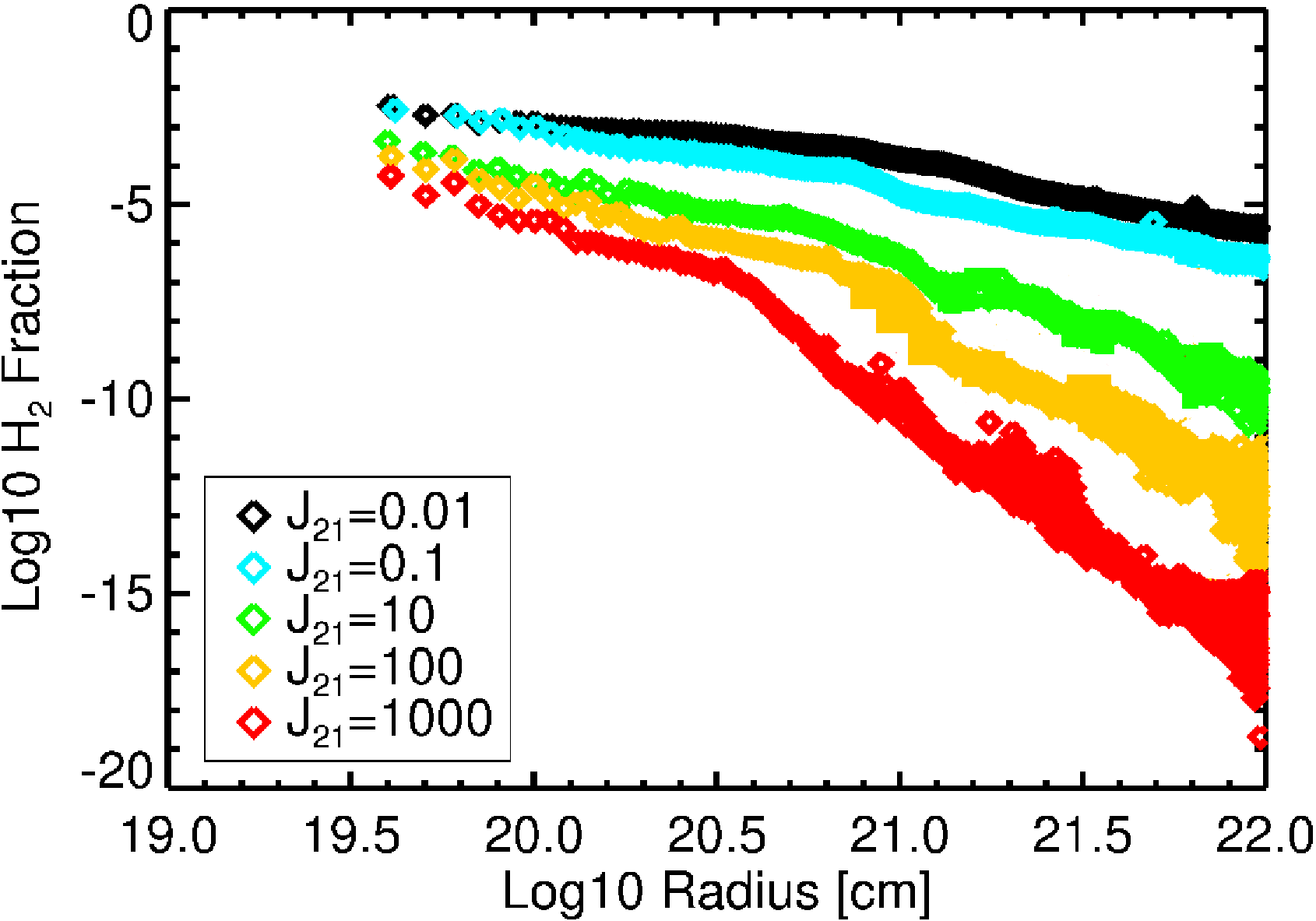}
\end{minipage} &

\begin{minipage}{8cm}
\includegraphics[scale=0.28]{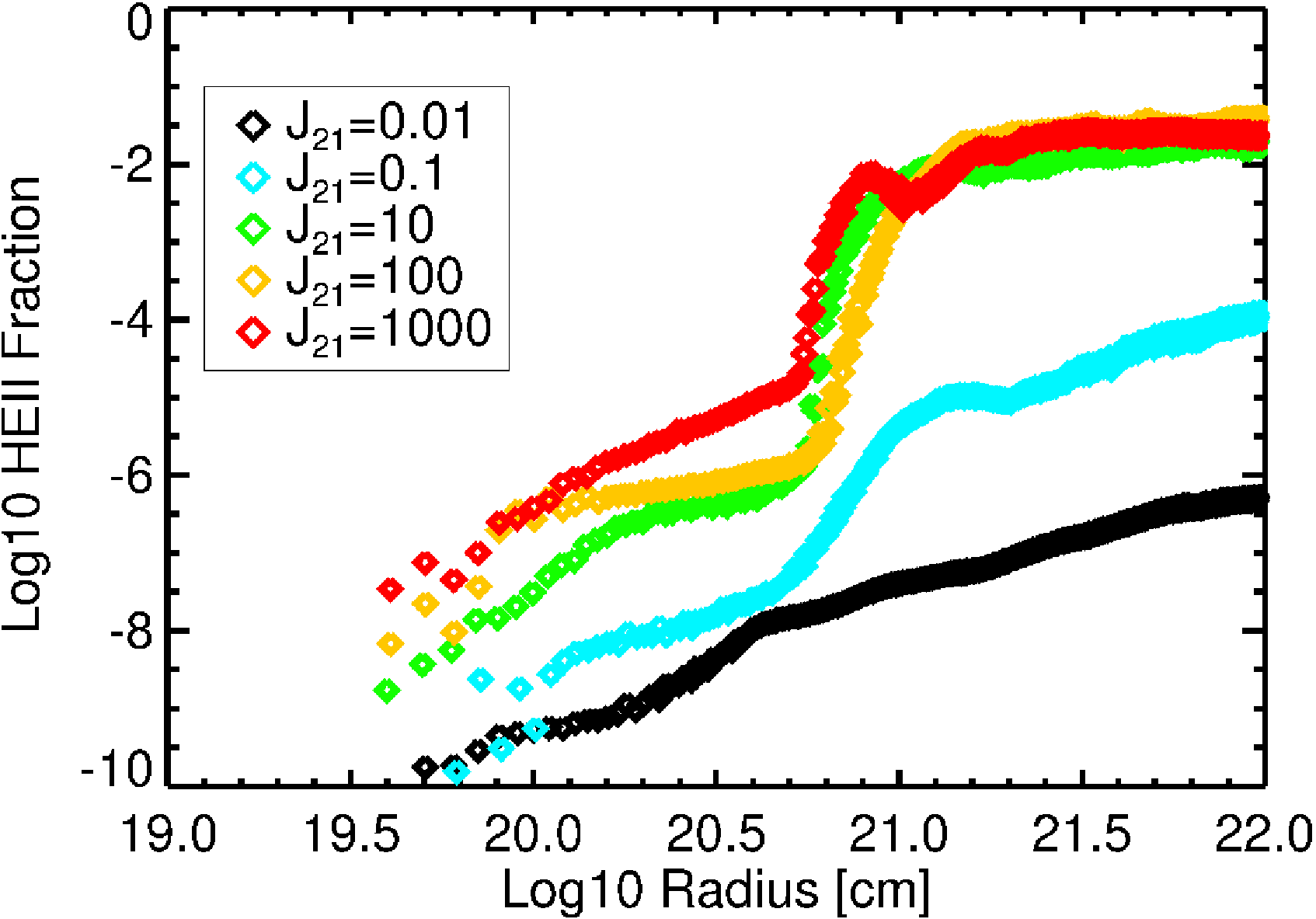}
\end{minipage}

\end{tabular}
\caption{ The escape fraction of ionizing radiation is 1\% for all the panels shown. Colors show different values of background intensities $\rm J_{21}$ as indicated in the legend. The upper left panel of this figure shows the temperature radial profile of the halo. The HII abundance radial profile for the halo is depicted in the upper right panel. $\rm H_{2}$ abundance is shown in the lower left panel. The HeII radial profile of the halo is depicted in the lower right panel. It can be noted that for higher values of $\rm J_{21}$, the IGM gas is heated up to $\rm 10^{5}$ K and the degree of ionization is enhanced. $\rm H_{2}$ is photo-dissociated for stronger radiation fields.}
\label{figure1}
\end{figure*}

\begin{figure*}[htb!]
\centering
\begin{tabular}{c c}
\begin{minipage}{8cm}
\includegraphics[scale=0.28]{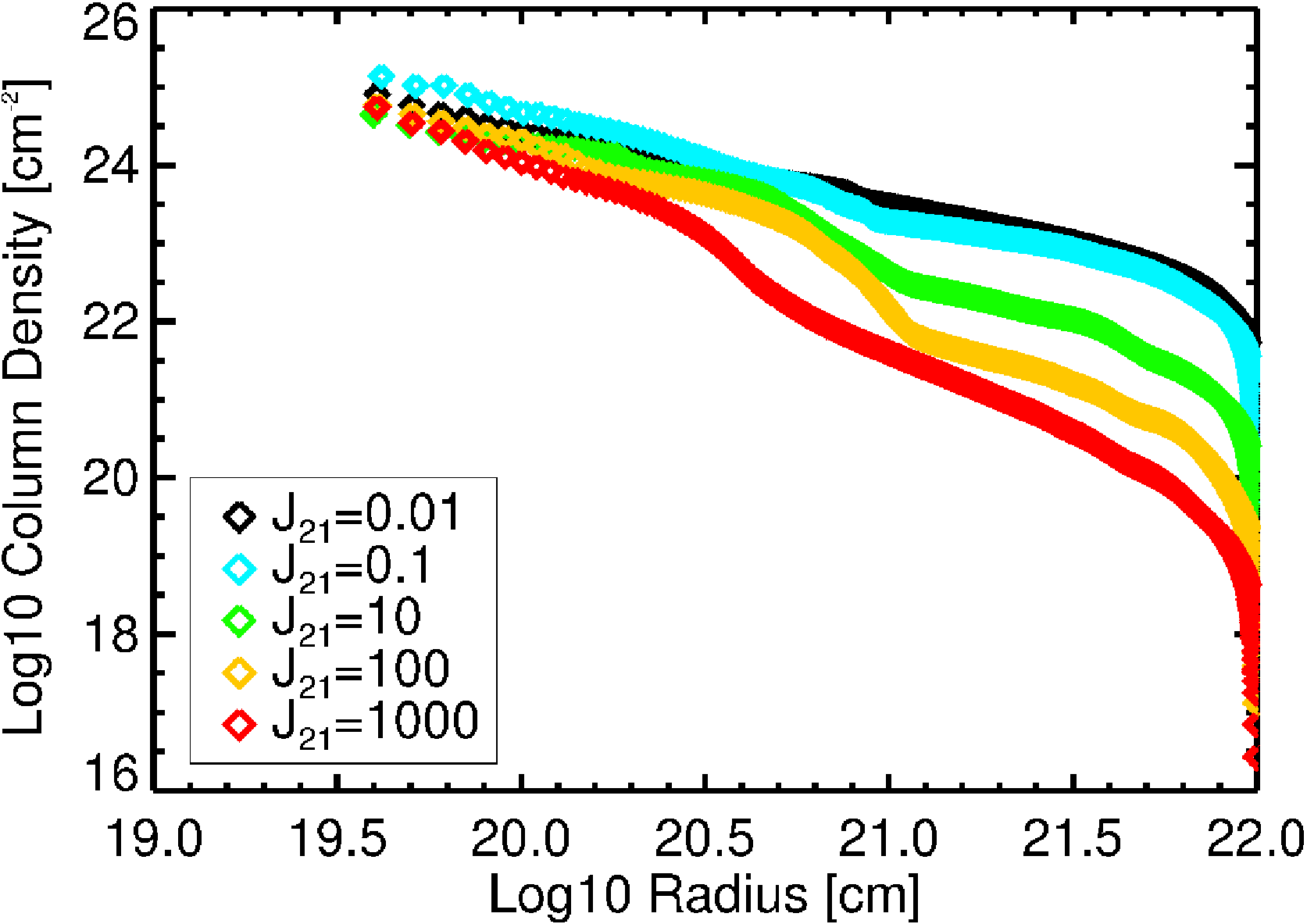}
\end{minipage} &
\begin{minipage}{8cm}
\includegraphics[scale=0.28]{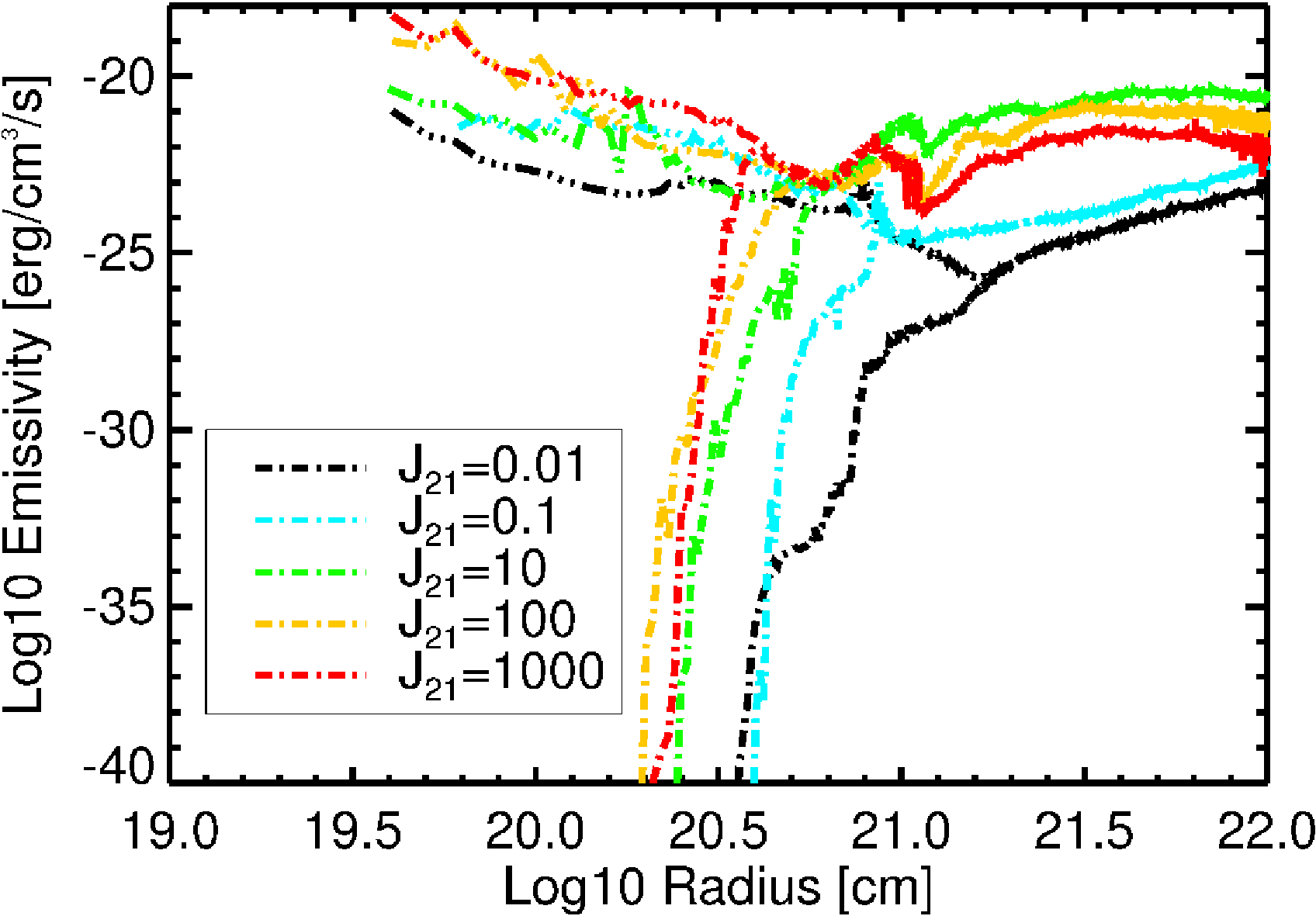}
\end{minipage} \\  \\

\begin{minipage}{8cm}
\includegraphics[scale=0.28]{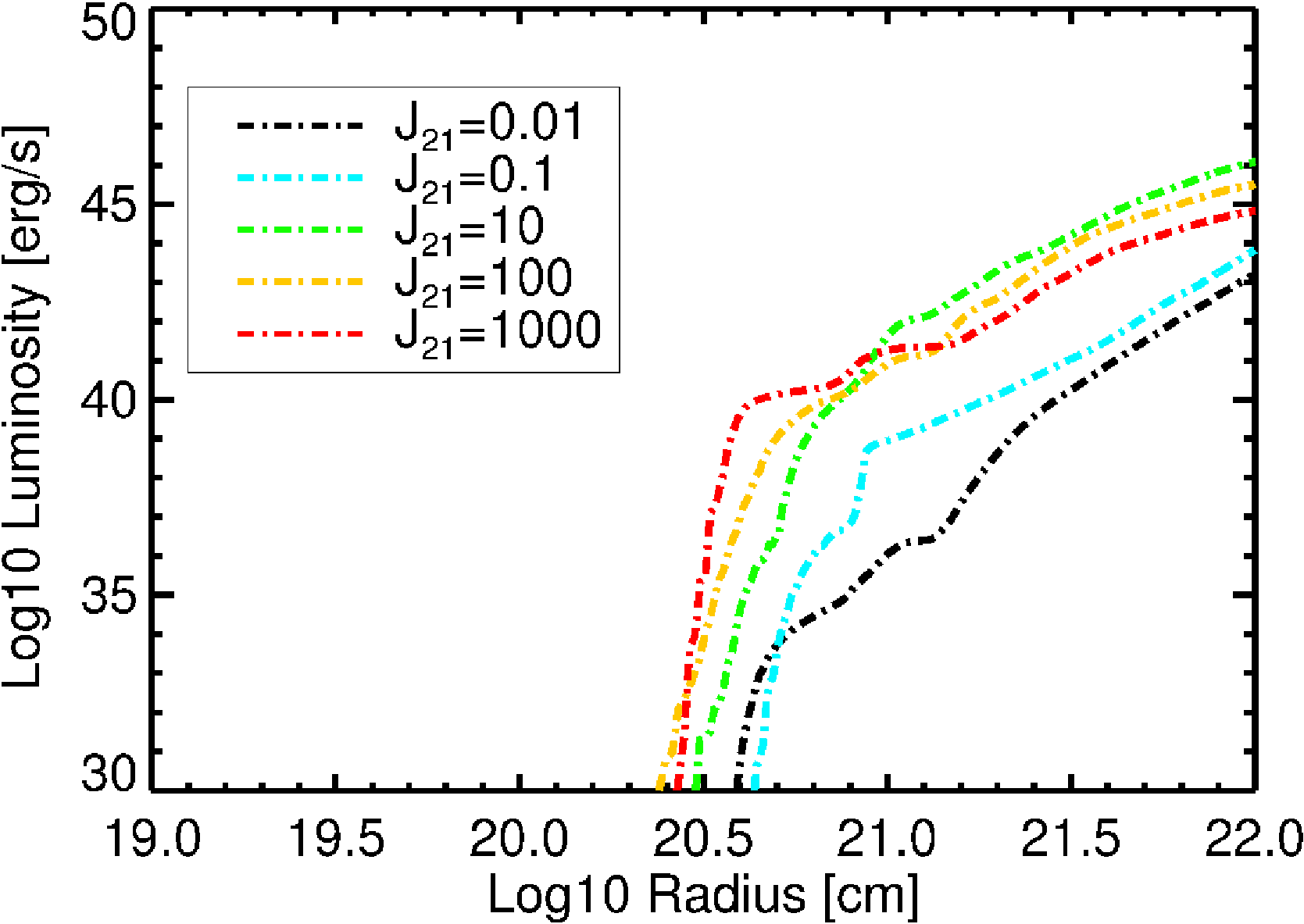}
\end{minipage} &

\begin{minipage}{8cm}
 \includegraphics[scale=0.28]{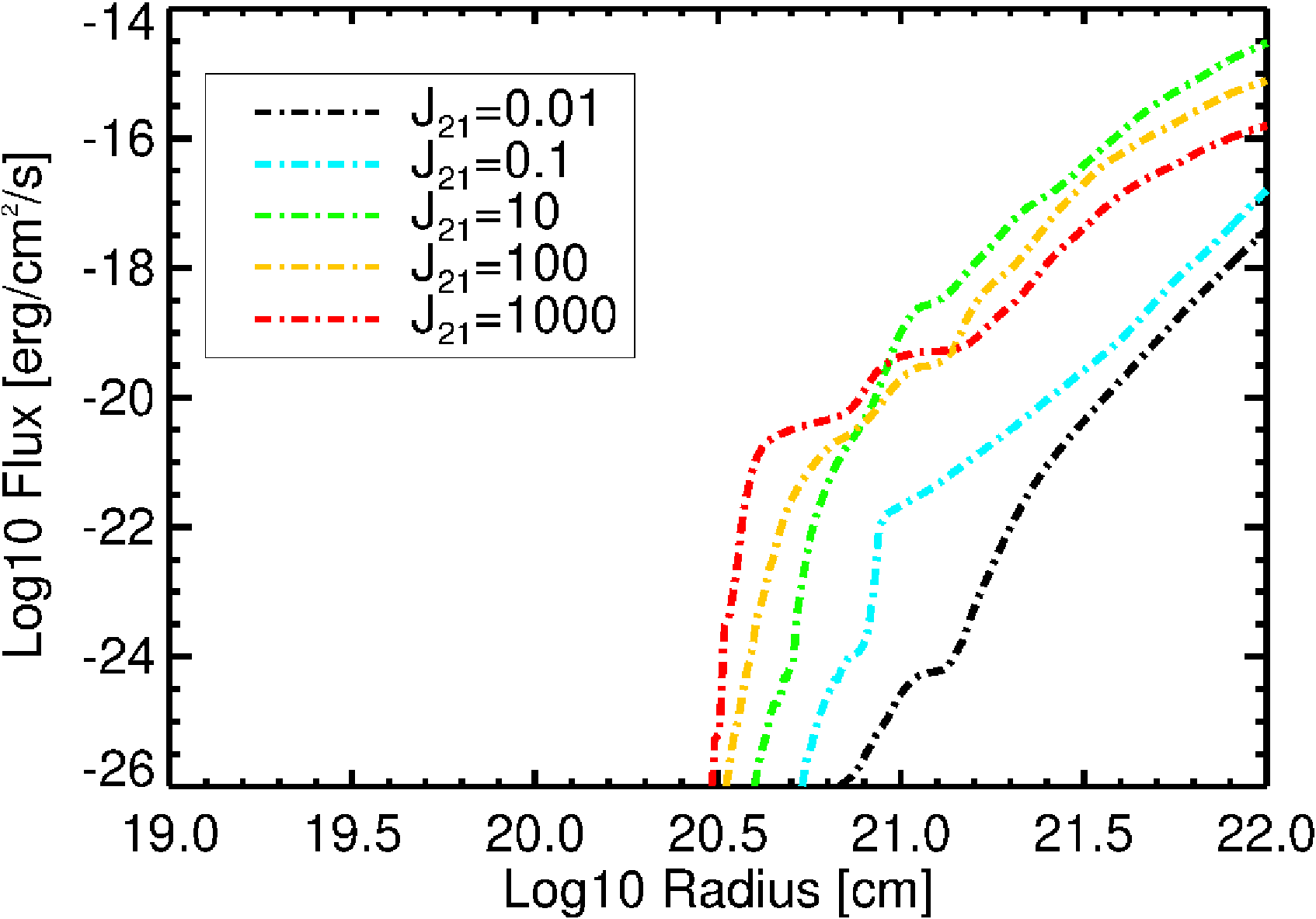}
\end{minipage}

\end{tabular}
\caption{The upper left panel shows the column density radial profile of the halo. The emissivity radial profile for the halo is depicted in the upper right panel, where the dot-dashed line shows the Lyman alpha emissivity and the triple dot-dashed line shows the total emissivity of all the coolants. The lower left panel shows the luminosity radial profile. The flux radial profile of the halo is depicted in the lower right panel. All panels in this figure have an escape fraction of 1\% for ionizing photons. Colors represent the different values of background intensity $\rm J_{21}$ as shown in the legend. Emissivity of Lyman alpha photons increases upto $\rm J_{21}=10$ and starts to decrease for intenser radiation fields as the neutral fraction goes down. Lyman alpha luminosity and flux show the same behavior. }
\label{figure2}
\end{figure*}


%
%
%

\begin{figure*}[htb!]
\centering
\begin{tabular}{c c}
\begin{minipage}{8cm}
\hspace{0.27cm}
\includegraphics[scale=0.28]{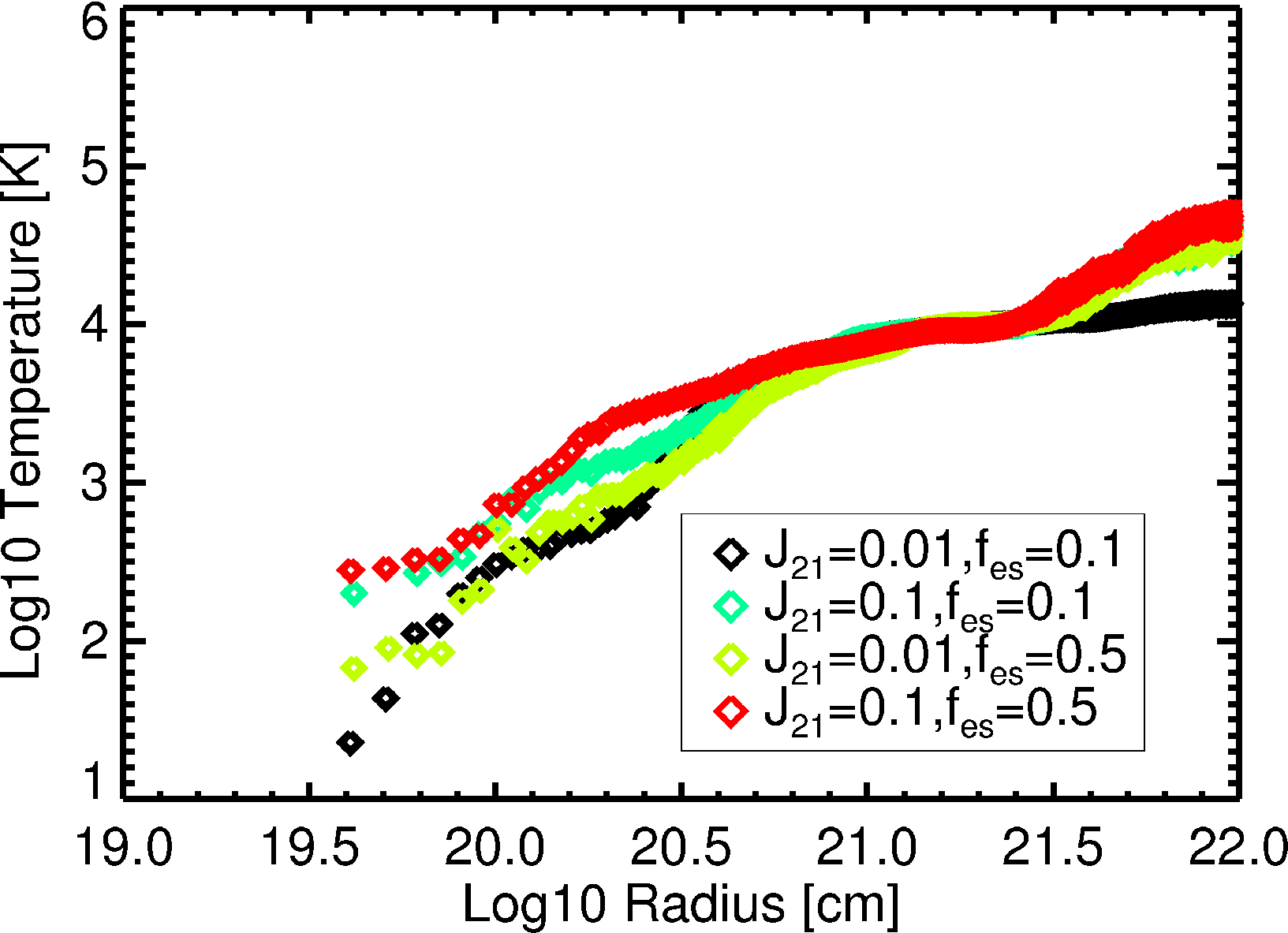}
\end{minipage} &
\begin{minipage}{8cm}
\includegraphics[scale=0.28]{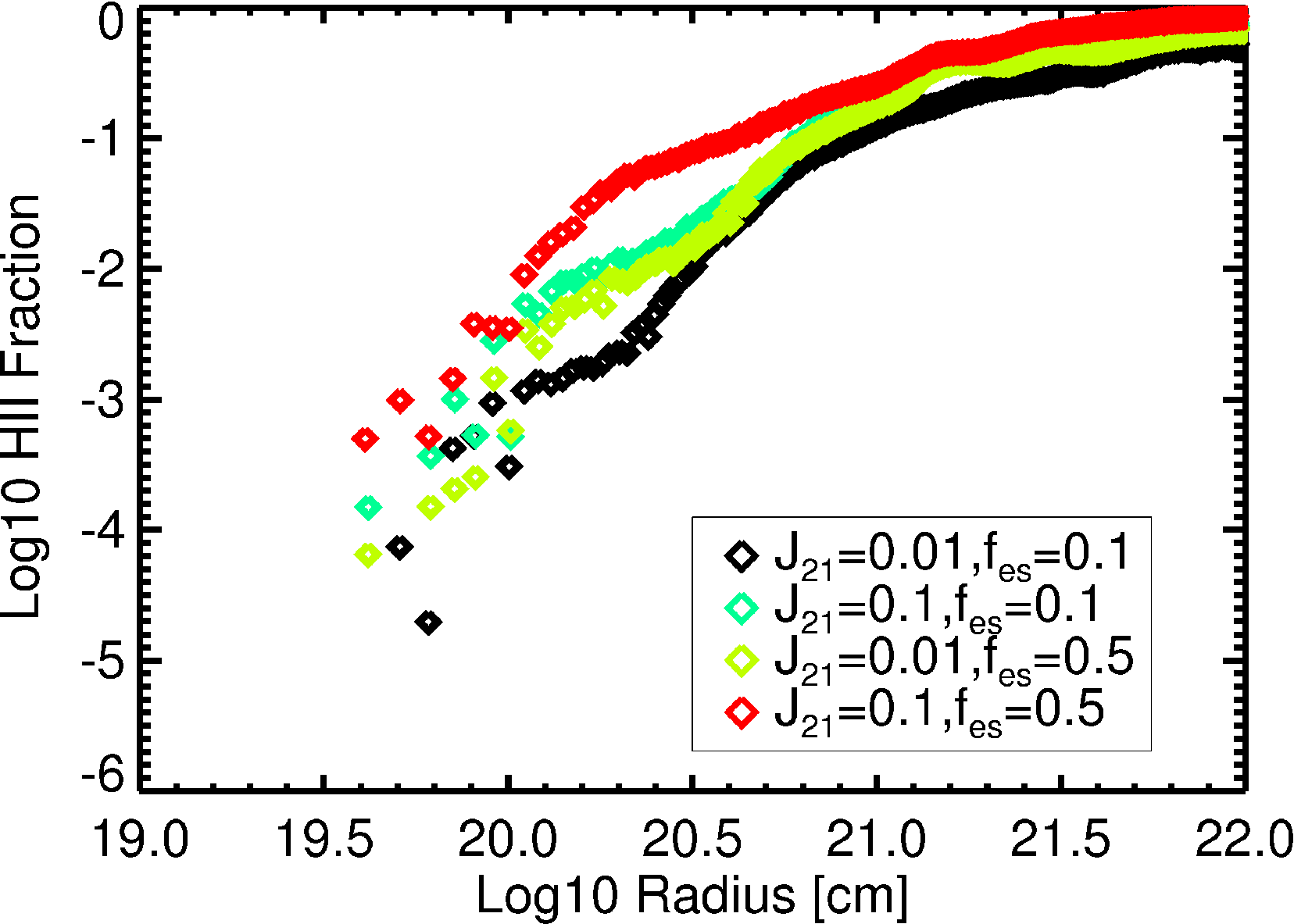}
\end{minipage} \\  \\

\begin{minipage}{8cm}
\includegraphics[scale=0.28]{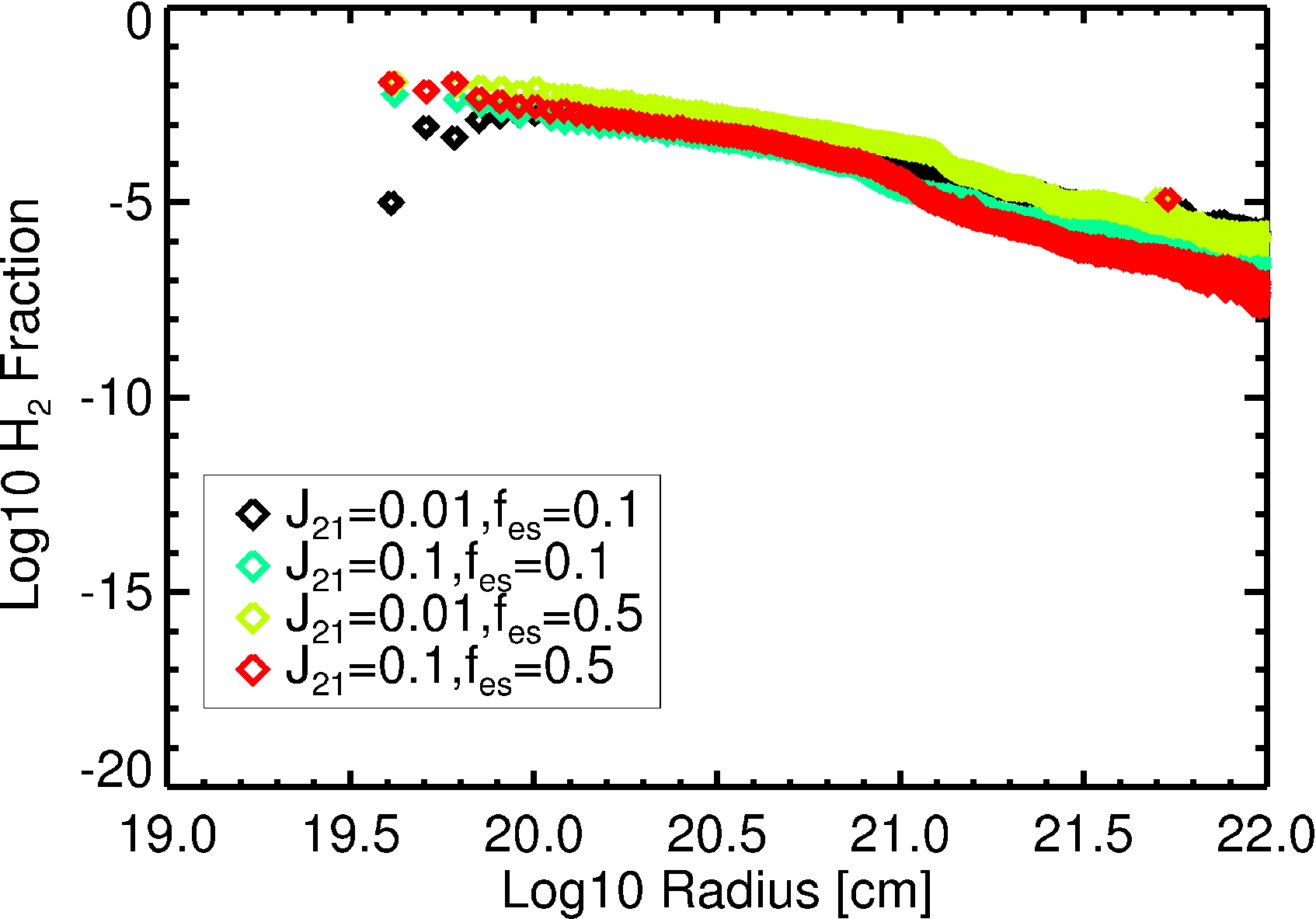}
\end{minipage} &

\begin{minipage}{8cm}
\includegraphics[scale=0.28]{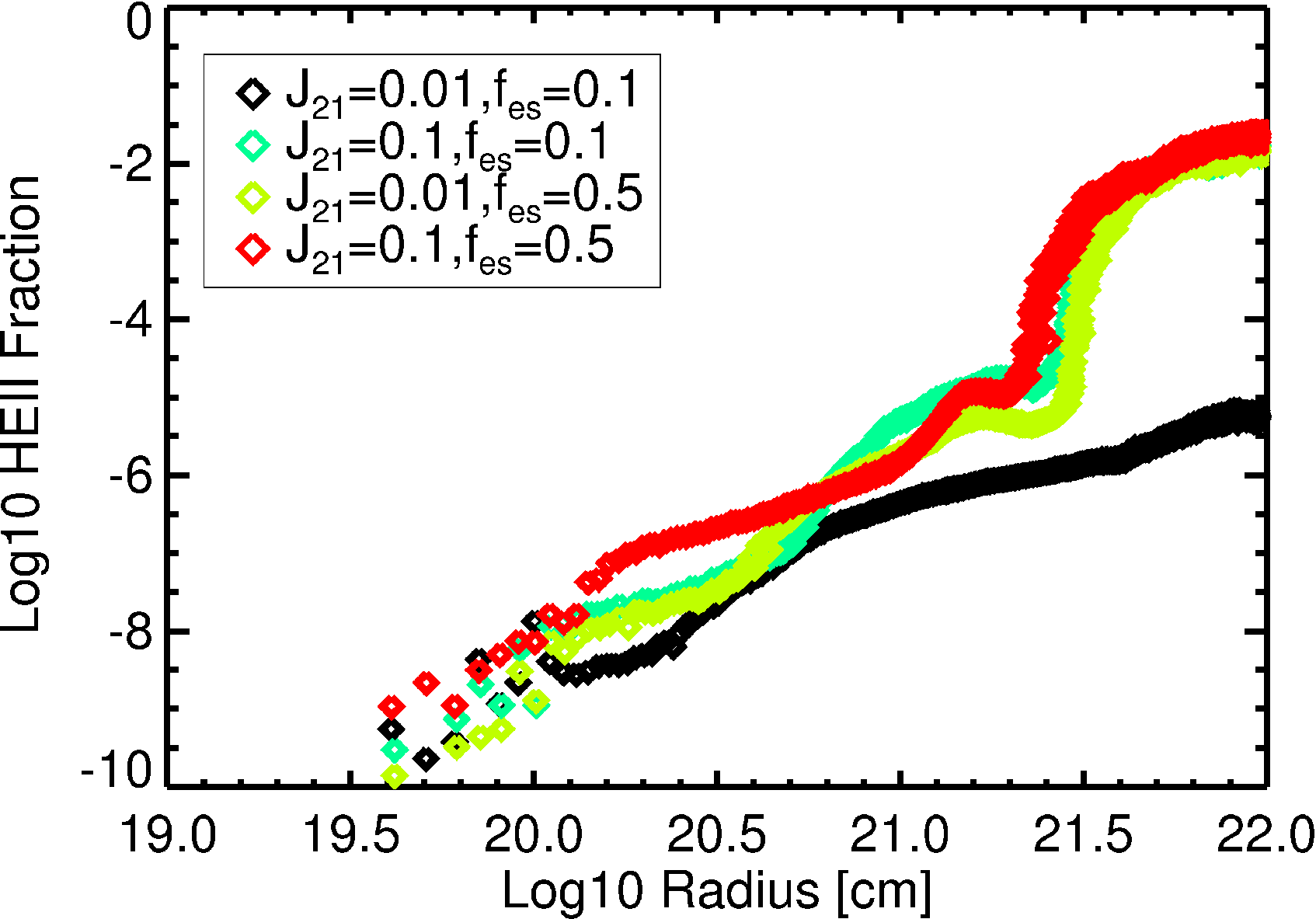}
\end{minipage}

\end{tabular}
\caption{ The upper left panel of this figure shows the temperature radial profile of the halo. The HII abundance radial profile for the halo is depicted in the upper right panel. $\rm H_{2}$ abundance is shown in the lower left panel. The HeII radial profile of the halo is depicted in the lower right panel. The legend in each panel shows the values of the escape fraction of ionizing photons. The corresponding strength of the background radiation field is also shown in the legend. It is found that using a higher escape fraction of ionizing radiation, the degree of ionization is enhanced and gas is heated upto higher temperatures for the same background radiation field strengths. The formation of $\rm H_{2}$ is not affected much by a higher escape fraction of ionizing photons.}
\label{figure5}
\end{figure*}

\begin{figure*}[htb!]
\centering
\begin{tabular}{c c}
\begin{minipage}{8cm}
\includegraphics[scale=0.28]{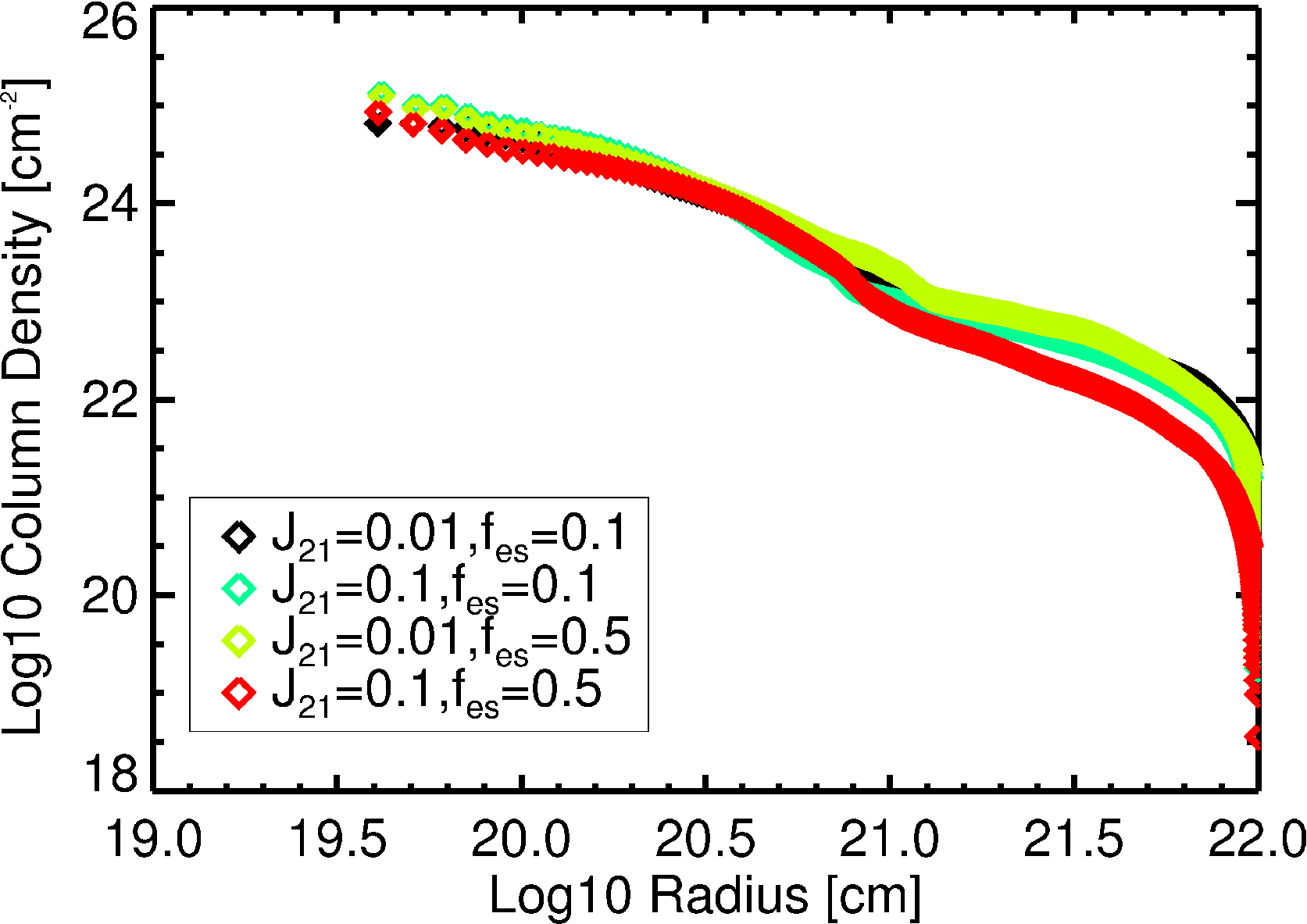}
\end{minipage} &
\begin{minipage}{8cm}
\includegraphics[scale=0.28]{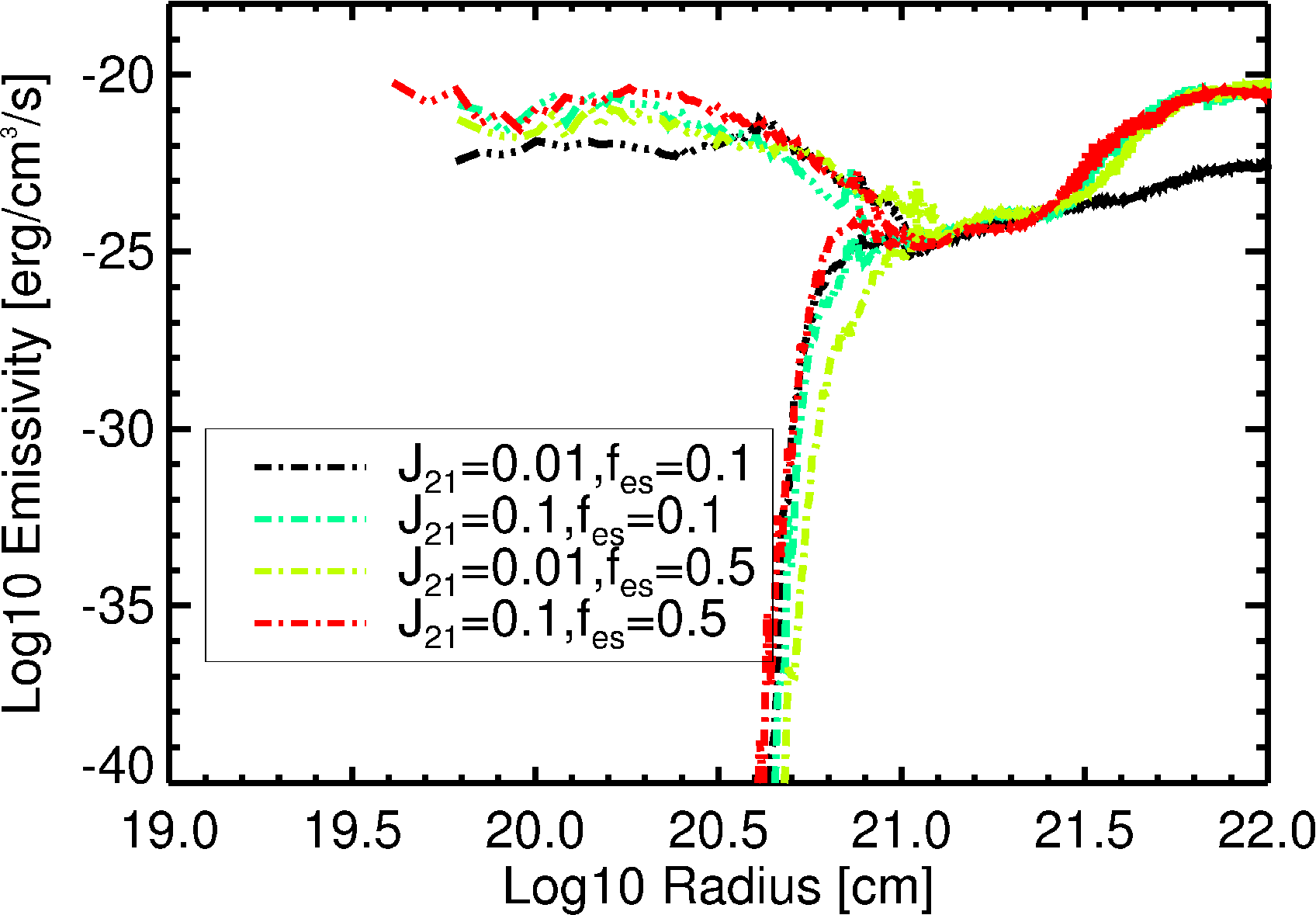}
\end{minipage} \\  \\

\begin{minipage}{8cm}
\includegraphics[scale=0.28]{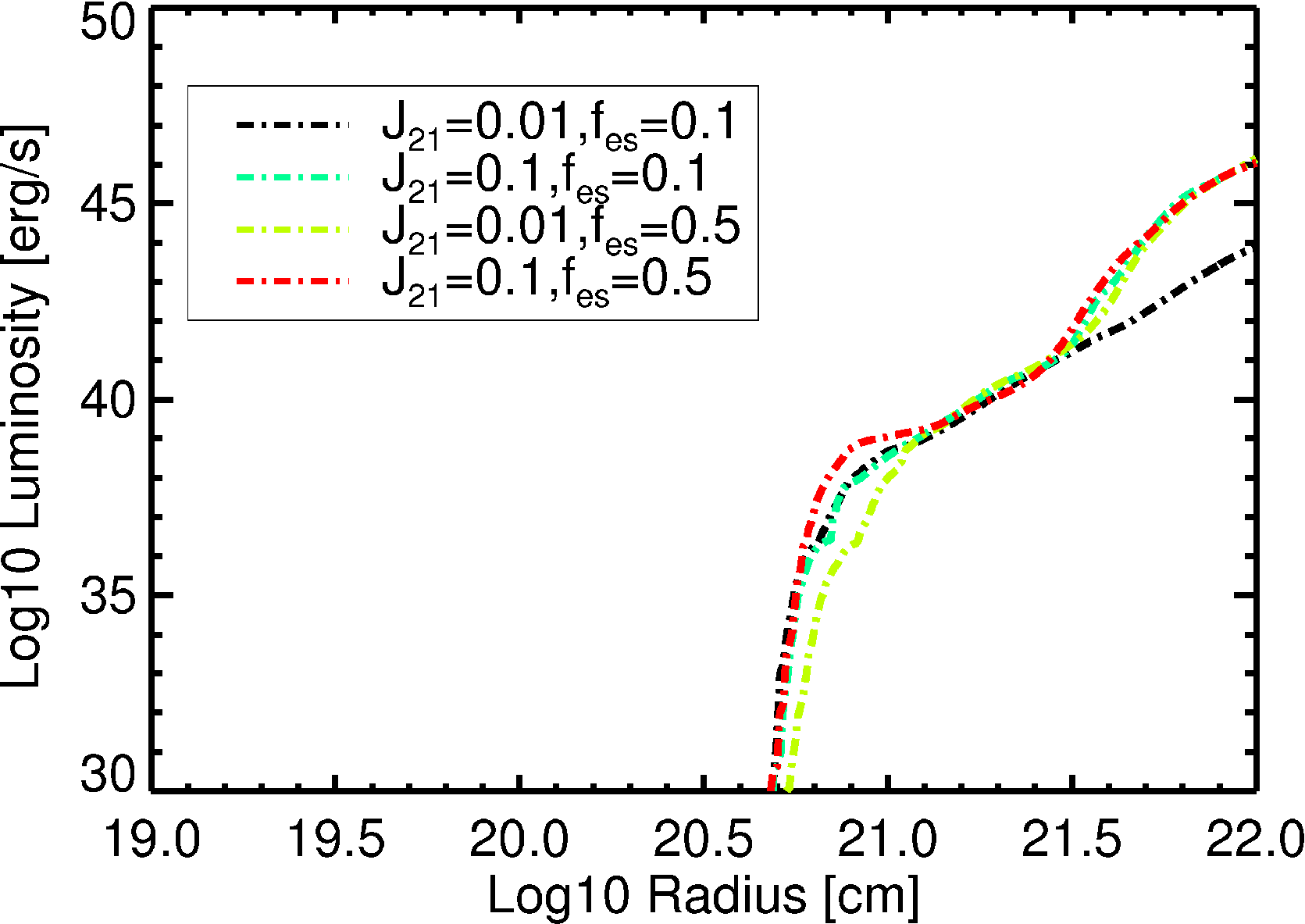}
\end{minipage} &

\begin{minipage}{8cm}
\includegraphics[scale=0.28]{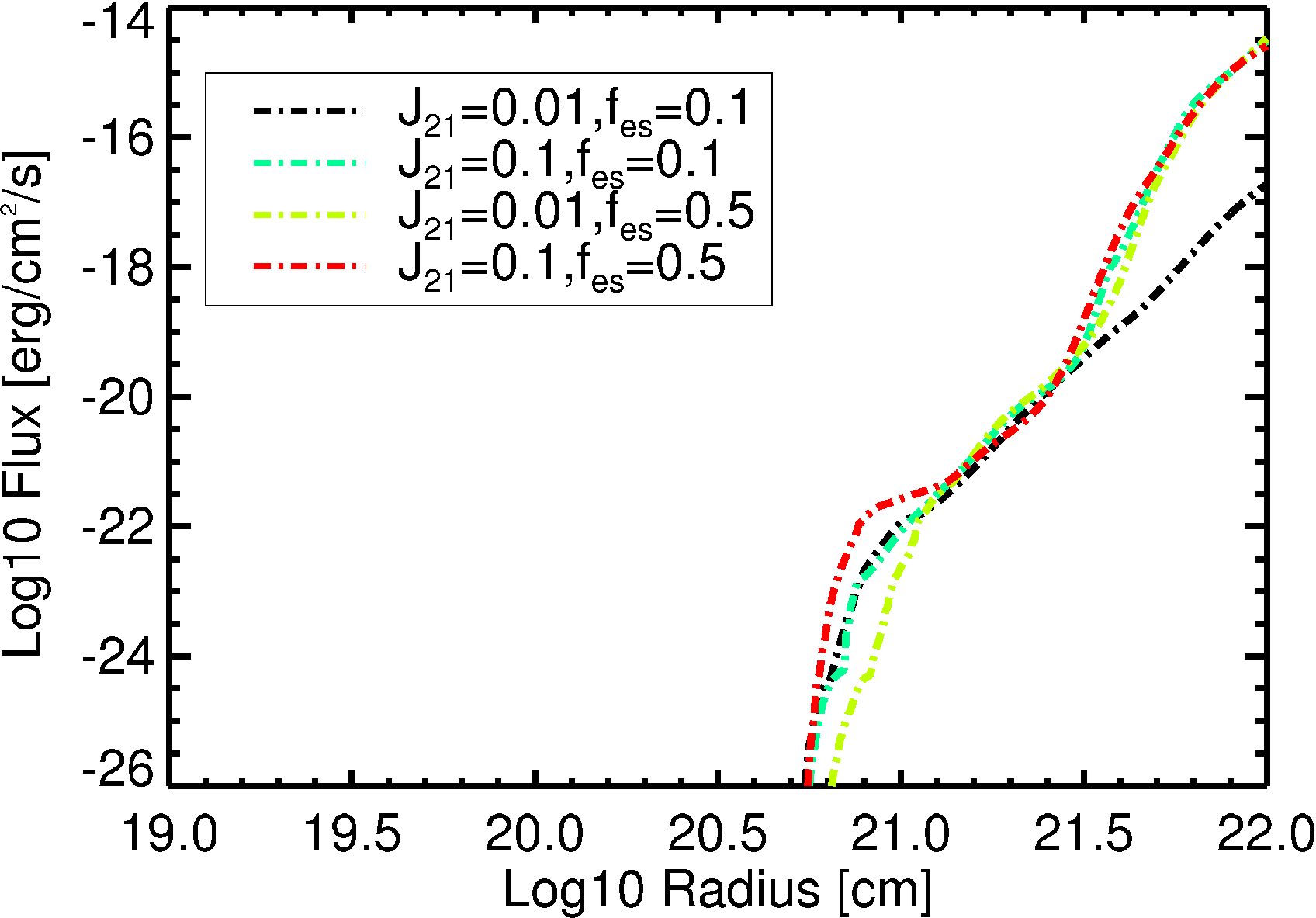}

\end{minipage}

\end{tabular}
\caption{The upper left panel shows the column density radial profile of the halo. The emissivity radial profile for the halo is depicted in the upper right panel, where the dot-dashed line shows the Lyman alpha emissivity while the triple dot-dashed line shows the total emissivity of all the coolants. The lower left panel shows the luminosity radial profile. The radial profile of the enclosed flux from the halo is depicted in the lower right panel. The escape fraction of ionizing radiation and the strength of background radiation field is shown in the legend of each panel. The emissivity of Lyman alpha is significantly enhanced for higher escape fraction and a given background UV field strength. The same holds for luminosities and fluxes.}
\label{figure6}
\end{figure*}

\begin{figure*}[htb!]
\centering
\begin{tabular}{c c}
\begin{minipage}{8cm}
\includegraphics[scale=0.28]{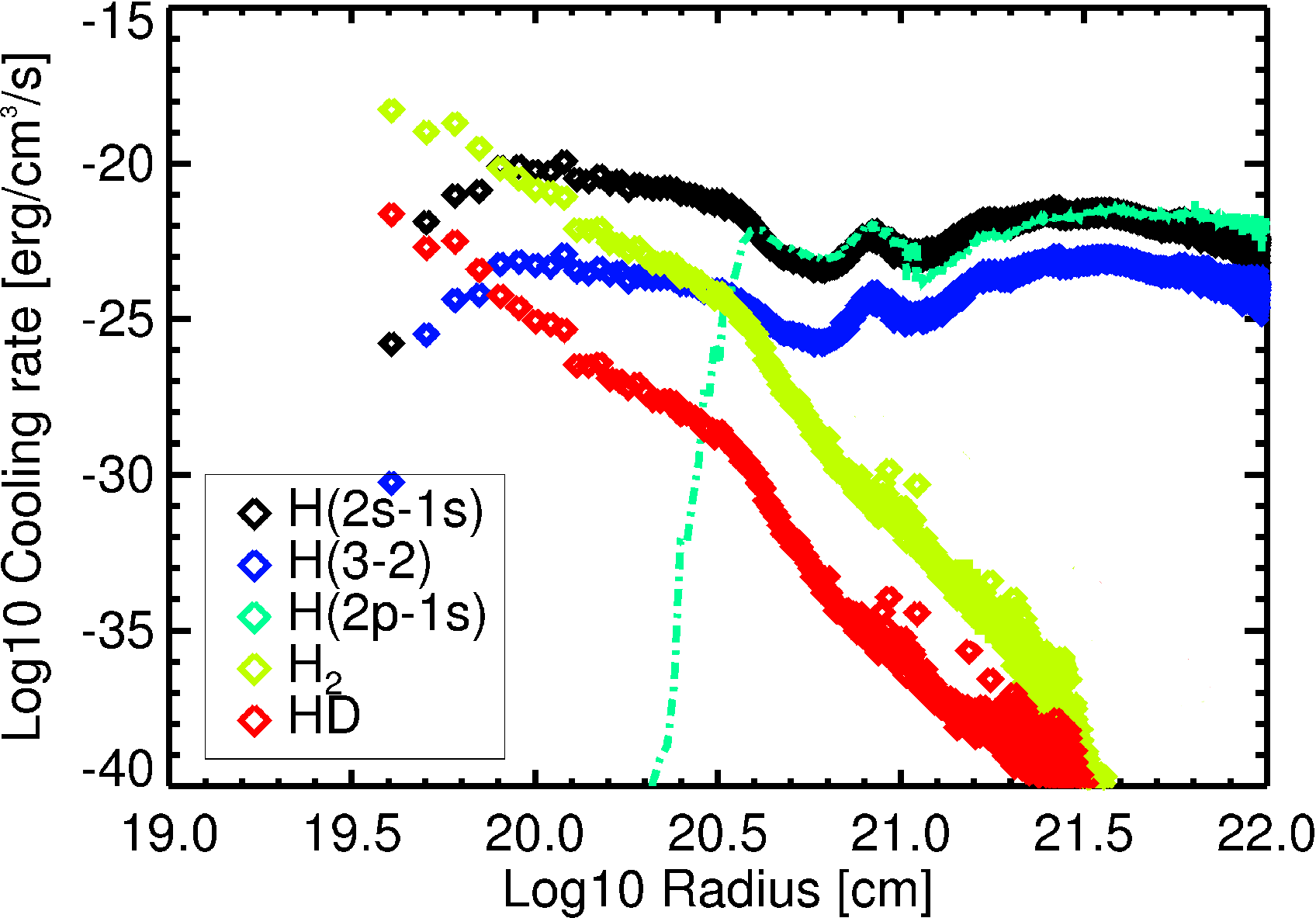}
\end{minipage} &
\begin{minipage}{8cm}
\includegraphics[scale=0.28]{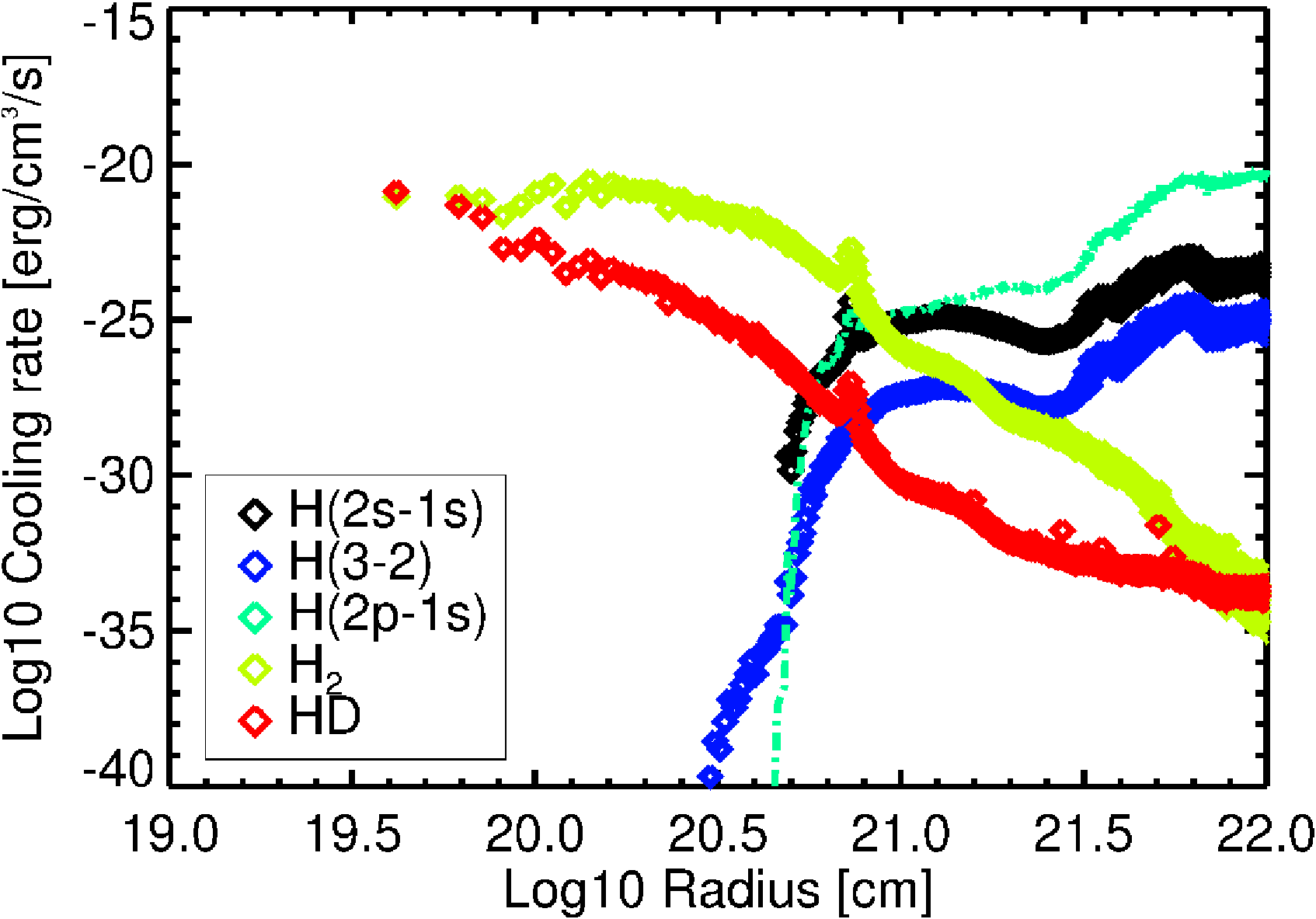}
\end{minipage} 
\end{tabular}
\caption{Cooling rates of $\rm H_{2}$, HD, as well as H(2p-1s), H(2s-1s) and H(3-2) transitions are plotted against radius in both panels. Left panel is for $\rm J_{21}=1000$ with a 1\% escape fraction of ionizing radiation. Right panel is for $\rm J_{21}=0.1$ with an escape fraction of 10\%.}
\label{h2hd}
\end{figure*}

\section{Results}


We perform nine cosmological simulations with different strengths of the background UV field and escape fractions of ionization radiation as listed in table \ref{tab:sumary}. We start our simulations at redshift 100. We note that density perturbations decouple from the Hubble flow and begin to collapse through gravitational instabilities. During the non-linear evolution phase gas collapses in the dark matter potentials and gets shock heated. We select a halo of $\rm 10^{9}~M_{\odot}$ at redshift 10 and follow its collapse down to redshift 5.5. At a redshift of about 8, gas becomes photo-ionized as we switch on the ionizing background produced by reionization. The temperature of the gas is raised above $\rm 10^{4}$ K due to photo-ionization heating depending upon the strength of the background flux. Photo-ionized gas recombines and begins to cool and collapse because of recombination cooling. The virialization process continuously transforms the gravitational potential energy into kinetic energy of gas and dark matter. Due to the dissipative nature of the gas, part of the gravitational energy goes into thermal energy and is radiated away. During virialization, the gas becomes turbulent and settles in the center of a halo through cold streams as shown in figure \ref{figure3}. The temperature of the gas in filamentary accretion is $\rm \sim 10^{4}$ K and densities are $\rm 10^{-2} -1~cm^{-3}$. The density radial profiles are shown in figure \ref{figure4}. It can be seen from the figure that density varies with distance r like $r^{-2.8}$. Our simulations reach a maximum density of a few times $\rm 10^{5}~cm^{-3}$. Small fluctuations in the density radial profiles are due to the clumpiness of the medium as the halo is not perfectly symmetric.
We show the radial velocity profile for three selected cases in the left panel of figure \ref{figvel}. The figure shows that gas is falling into the center of the halo. The variations in the radial velocity profile at the outer radii are due to different accretion rates for various background radiation fields. The gas in the center of the halo becomes stable. The right panel of figure \ref{figvel} shows the tangential velocity radial profile. Variations in tangential velocity are caused by shocks. The total rms velocity hardly depends on the radiation field, as it is mostly due to the virial temperature. We find typical velocities of about $\rm \sim 50~km/s$ as expected from the virial temperature. Virialization shocks also play an important role in the emission of Lyman alpha, for further details see \cite{2011MNRAS.tmpL.217L}.

We see that for $\rm J_{21} < 10$, the temperature of the gas is $\rm \sim 10^{4}$ K while for higher intensities (i.e. $\rm J_{21} > 10$) it reaches $\rm 10^{5}$ K. The temperature of $\rm 10^{5}$ K is driven by the ionization of helium, which absorbs more energetic photons. Cooling due to helium lines becomes effective at temperatures above  $\rm 5 \times 10^{4}$K and keeps the gas temperature at $\rm \sim 10^{5}$ K. At densities of 1 $\rm cm^{-3}$, cooling due to radiative recombination becomes very effective and cools the gas down to $\rm 10^{4}$ K. The gas in the surroundings of the halo remains hot ($ \rm > 10^{4}~K$) as shown in the top left panel of figure \ref{figure1}. Similarly, the degree of ionization varies depending upon the impinging radiation intensity, as shown in the upper right panel of figure \ref{figure1}. It is small for lower fluxes and large for higher fluxes. We also see that helium remains mostly neutral for lower fluxes and becomes singly ionized for a stronger radiation background. The He$^+$ abundance is shown in the bottom right panel of figure \ref{figure1}. At densities $\rm > 1~cm^{-3}$, the recombination rate becomes high and the gas begins to recombine. Consequently, the degree of ionization declines.

For low background fluxes, a  $\rm H_{2}$ abundance of $\rm 10^{-6}$ is formed at low densities (i.e. $\rm 10^{-2}~ cm^{-3}$) and a temperature of $\rm 10^{4}$ K, while its formation remains inhibited for higher fluxes at low densities. For a high background radiation field, i.e., $\rm J_{21} \geq 1000$, $\rm H_{2}$ is photo-dissociated even in the center of the halo. For weaker radiation fields photo-dissociation of molecular hydrogen becomes ineffective and the $\rm H_{2}$ abundance reaches the universal value (i.e., $\rm 10^{-3}$). The $\rm H_{2}$ radial profile for different fluxes is shown in the bottom left panel of figure \ref{figure1}. Our results are consistent with previous studies \citep{2010MNRAS.402.1249S,2008MNRAS.391.1961D,2003ApJ...596...34B,2011MNRAS.410..919J}. We also computed the column densities for different background fluxes, finding that the column density is higher for low fluxes in the envelope of a halo due to a higher neutral gas fraction. Conversely, the column is low for higher radiation field strengths because of a higher ionization degree. The upper left panel of figure \ref{figure2} shows the column density radial profile.

Based on these results, we computed the emissivity of Lyman alpha photons, finding that it is enhanced in the presence of a background incoming radiation field. It increases with the strength of the radiation field up to $\rm 10^{-22}~erg/cm^{2}/s/Hz/sr$ (i.e. $\rm J_{21} \times f_{esc}$) and then starts to decline for higher intensity. This decline in emissivity is a consequence of the decreasing abundance of atomic hydrogen. At columns higher than $\rm 10^{22}~cm^{-2}$, trapping of Lyman alpha becomes effective and consequently the emission of Lyman alpha photons diminishes while cooling still proceeds through higher electronic transitions of hydrogen. For a fixed ionizing background, the emission increases with increasing photo-dissociation flux. The emissivity of Lyman alpha photons as well as the total emissivity is plotted  against radius in the upper right panel of Fig. \ref{figure2}. Lyman alpha emission is higher in the envelope of the halo and sharply declines towards the center. This behavior of Lyman alpha emission is due to line trapping of Lyman alpha photons and is consistent with previous studies \citep{2011MNRAS.tmpL.217L,2010ApJ...712L..69S,2006ApJ...652..902S}. In addition to the line trapping, also the cooling by $\rm H_{2}$ and HD may prevent Lyman alpha emission, as it drives the gas to lower temperatures (i.e. the run $\rm J_{21}=10$). The total emissivity remains roughly constant as cooling is compensated by two-photon continuum processes. Moreover, for higher background radiation field strengths the total emissivity from hydrogen electronic states is higher than for low radiation fields. This is due to a higher molecular hydrogen fraction for low fluxes and a lower $\rm H_{2}$ fraction for high fluxes through photodissociation. The local variations in emissivity are because of the clumpiness of the medium.


The enclosed Lyman alpha luminosity as a function of radius is shown in the bottom left panel of figure \ref{figure2}. The luminosity increases with the strength of the ionizing radiation field up to 0.1 (in units of $\rm J_{21}$) and then decreases with increasing radiation field strength similar to the emissivity. Luminosity values are of the order of $\rm 10^{43}-10^{46}~erg/s$ which are comparable to observed Lyman alpha blobs as well as estimates from numerical simulations \citep{2009ApJ...696.1164O,2010Natur.467..940L,2010MNRAS.407..613G,2006ApJ...649...14D}. We also computed the flux. We found that the flux mostly emerges from the envelope of a halo rather than its center and is significantly enhanced in the presence of a background UV field. The total flux varies from $\rm 10^{-18}-10^{-15}~ erg/cm^{2}/s$ depending upon the ambient radiation field strength. Our results are in agreement with our previous finding that emission of Lyman alpha is extended and emerges from the hot gas in the surroundings of a halo.

Due to large uncertainties in the escape fraction of ionizing radiation \citep{2010arXiv1006.3519F}, we have performed a number of simulations to assess the impact of different escape fractions on the emissivity of Lyman alpha photons. We performed simulations with escape fractions of 1\%, 10\% and 50 \% for a background radiation field of $\rm J_{21}=0.01~and~0.1$. We found that for a similar value of $J_{21}$ describing the strength of photodissociation, an increasing escape fraction of ionizing photons enhances the temperature of the gas. Radial profiles of temperature, H$^+$, He$^+$ and $\rm H_{2}$ abundances are shown in figure \ref{figure5}. As expected, the degree of ionization increases for higher escape fraction. The gas is still neutral in the center of the halo but becomes ionized in the surroundings. Column density is shown in the upper right panel of figure \ref{figure6}. It is similar to the case with low escape fraction. Differences in the columns is due to different gas ionization fractions. The upper left panel of figure \ref{figure6} shows the emissivity of Lyman alpha photons. It is found that for the same background value of $\rm J_{21}$, the emissivity is enhanced for higher escape fractions. It reaches a maximum in the envelope of the halo and sharply drops as trapping becomes effective in the center. The total emissivity is again compensated by cooling through higher states of hydrogen atoms. There are some local variations in the emissivity profile due to the density structure, fluctuations in temperature and the ionization degree. We also computed the luminosity and flux for this case and found that the flux is boosted for a higher escape fraction. The radial profiles of luminosity and flux are shown in the bottom panels of figure \ref{figure6}. Luminosities and fluxes for all cases are summarized in table \ref{tab:sumary}.

\begin{figure}[!htb]
\centering
\includegraphics[scale=0.28]{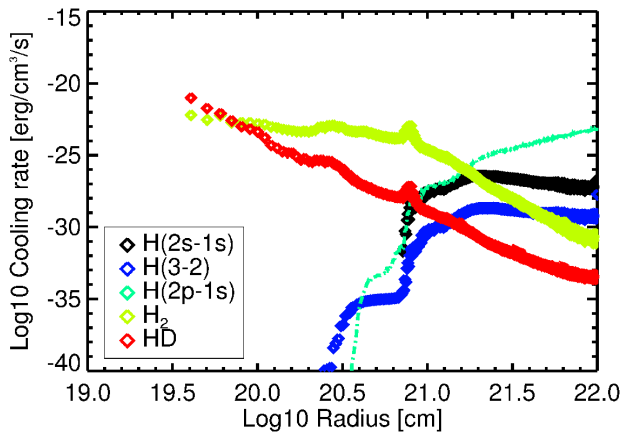}
\caption{Radial profiles of different coolants as mentioned in the legend for a background radiation field of $\rm J_{21}=0.01$ and a 1\% escape fraction of ionizing radiation.}
\label{h2hd1}
\end{figure}

We also computed the emissivity of various coolants to assess their relative importance for the cooling. We here show three typical cases for background UV fluxes of $\rm J_{21}=0.01,~0.1~and~1000$. Radial profiles of cooling rates from $\rm H(2p-1s),~H(2s-1s)~and~H(3-2)$ transitions as well as $\rm H_{2}~and~HD$ ro-vibrational transitions are shown in figures \ref{h2hd} and \ref{h2hd1}. The left panel of figure \ref{h2hd} shows the cooling rates for $\rm J_{21}=1000$. It is found that at radii $\rm > 3 \times 10^{20}$ cm cooling is only coming from electronic transitions of hydrogen. Notable transitions are H(2p-1s), H(2s-1s) and H(3-2). The population of these transitions is regulated by radiative de-excitation and collisional excitation, which makes them efficient coolants. The cooling from $\rm H_{2}~and~HD$ remains negligible at these radii. At radii between $\rm 3 \times 10^{20} -10^{20}$ cm line trapping of Lyman alpha photons becomes effective and two-photon continuum emission (2s-1s transition) becomes dominant. In the center of the halo ($\rm <10^{20}$ cm) cooling is regulated by trace amounts of molecular hydrogen. The right panel of figure \ref{h2hd} shows the cooling rates for $\rm J_{21}=10$. At radii $\rm > 10^{21}$cm  cooling is dominated by electronic transitions of hydrogen. Molecular hydrogen becomes an effective coolant for radii between $\rm 10^{21}-10^{20}$cm as cooling from electronic transitions of hydrogen becomes negligible due to lower gas temperatures. In the core of the halo, HD cooling dominates over $\rm H_{2}$ cooling for $\rm J_{21} \leq 0.1$. It can be seen in figure \ref{h2hd1} that, similar to previous cases, at radii $\rm > 10^{21}$ cm the halo is cooled by electronic transitions of hydrogen. Emission from these transitions becomes negligible at lower radii due to lower gas temperatures ($< 10^{3}$ K). Again, cooling in the center of the halo is dominated by HD.


\section{Discussion and Conclusions}

We have performed 9 simulations in total to investigate the role of the UV background radiation field in the emission of Lyman alpha photons. We used the adaptive mesh refinement code FLASH to carry out cosmological simulations and coupled it with a chemical network that solves the rate equations of 12 species. Our chemical network includes collisional ionization, radiative recombination, photo-ionization and photo-dissociation processes, the formation of molecules and a multi-level model of atomic hydrogen. We found that the presence of a background flux photo-ionizes the gas and raises the gas temperature to $\rm \geq 10^{4}$ K. We found that for weak radiation fields ($\rm J_{21}<100$),  cooling due to $\rm H_{2}$  decreases the temperature to at least 1000 K. For $\rm J_{21}<0.1$, cooling by HD is important as well, leading to gas temperatures of $\rm \sim 100$ K. We note that the HD cooling function was recently revisited by \cite{2011arXiv1103.2957C}. However, they only find differences at high temperatures, where $\rm H_2$ dominates the cooling. The thermal evolution in these cases is remarkably different and provides an environment conducive to star formation in the halos irradiated by weaker radiation fields, while  stronger fields may lead to the formation of massive objects.

We noted that the ubiquity of background flux enhances the emission of Lyman alpha photons. We found that the emission of Lyman alpha photons is boosted with ionizing flux up to $\rm 10^{-22}~erg/cm^{2}/s/Hz/sr$. The emission starts to decrease for higher values of the ionizing flux. This decline in the emissivity of Lyman alpha photons is because of the small abundance of atomic hydrogen in the presence of intense radiation fields. The flux predominantly arises from collisional excitation of atomic hydrogen. We found that the contribution from the recombination is at least few orders of magnitudes smaller.

\begin{figure*}[htb!]
\centering
\begin{tabular}{c c}
\begin{minipage}{8cm}
\includegraphics[scale=0.28]{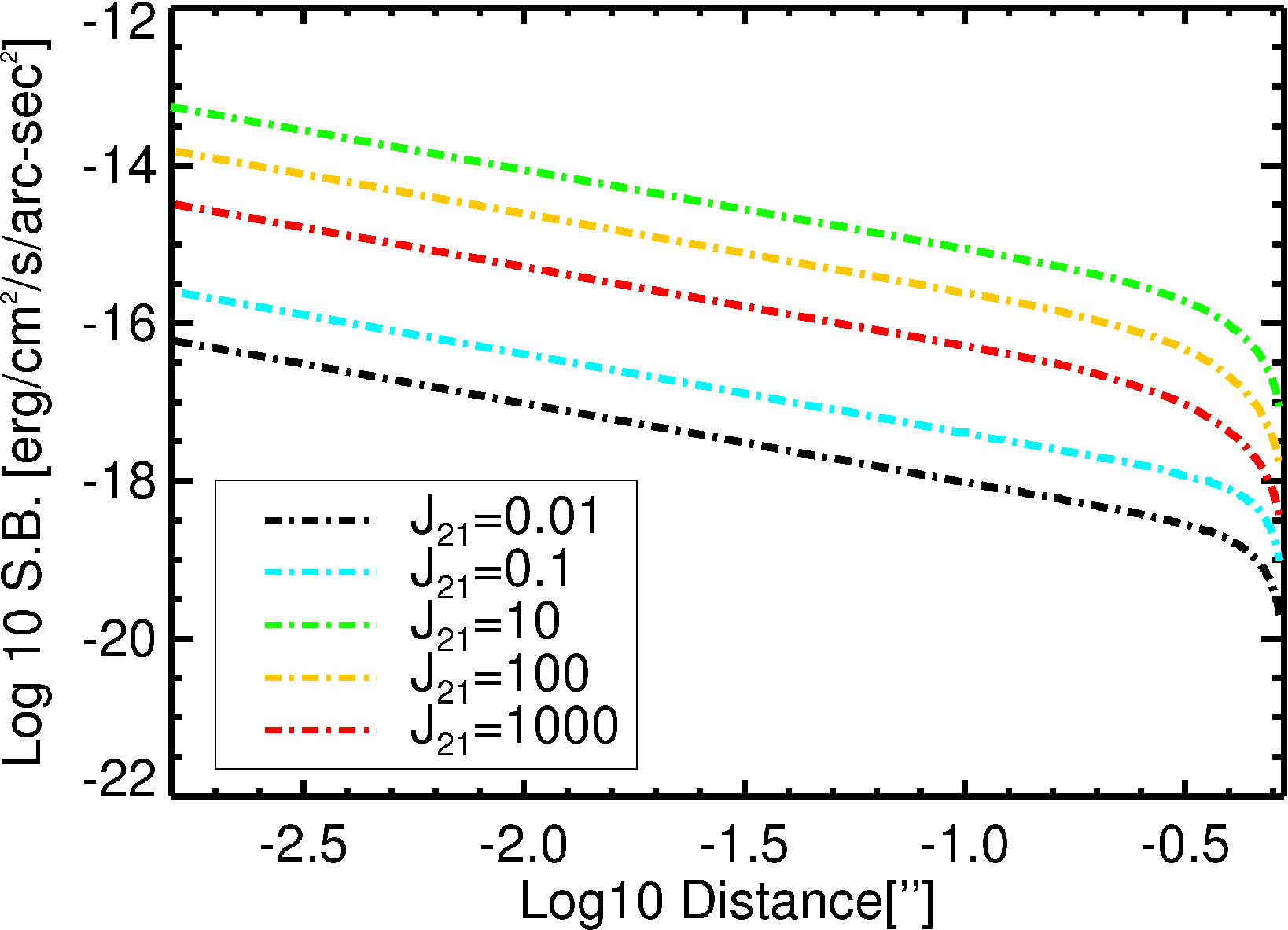}
\end{minipage} &
\begin{minipage}{8cm}
\includegraphics[scale=0.28]{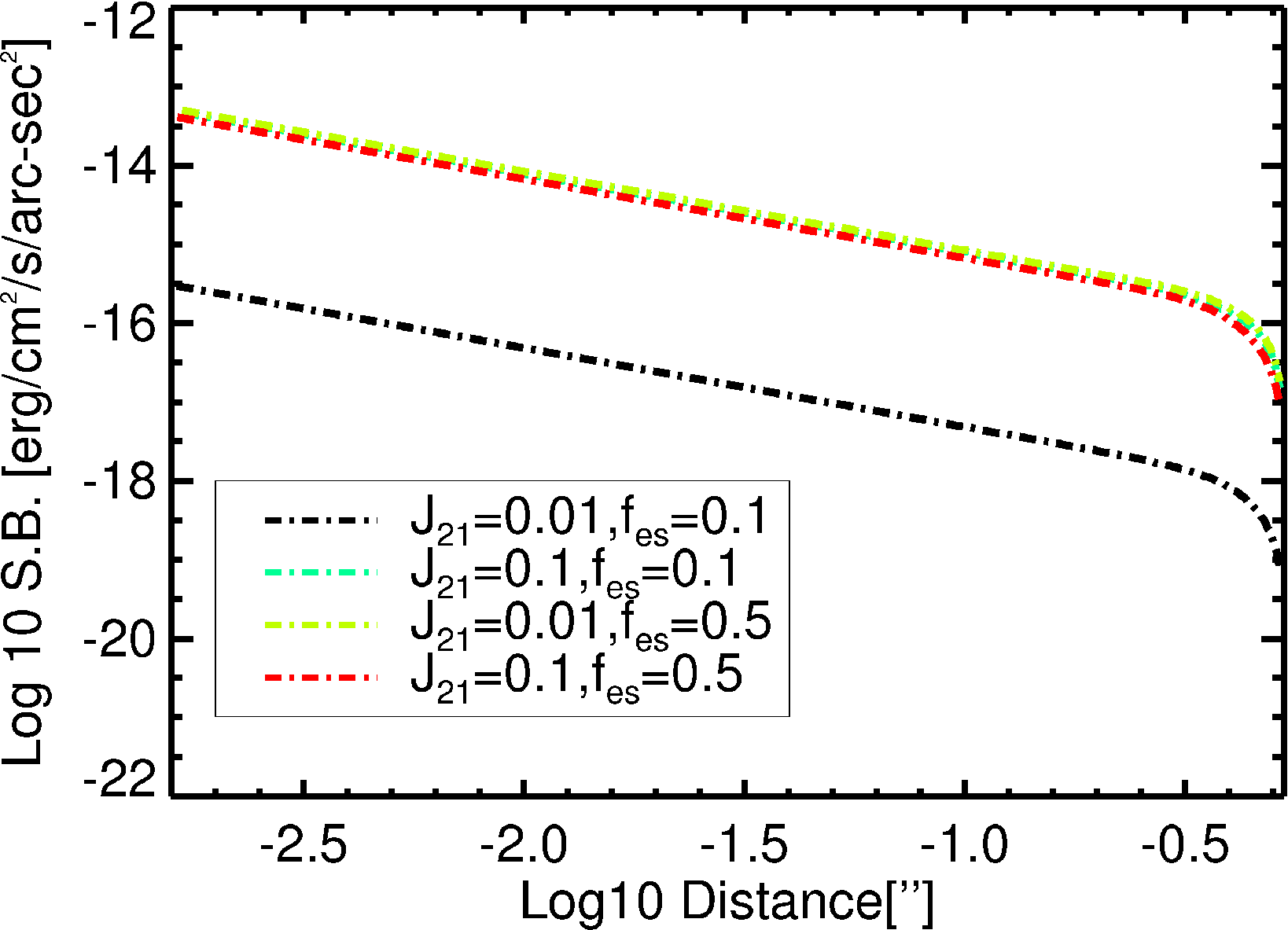}
\end{minipage} 
\end{tabular}
\caption{Both panels of this figure show the surface brightness profiles for various strengths of background UV radiation fields as mentioned in the legends. The value of the escape fraction is 1\% for the left panel. For the right panel, the value of the escape fraction against each radiation field is shown in the legend.}
\label{sbfig}
\end{figure*}

We also noted that the emission of Lyman alpha photons emerges from the envelope of a halo rather than its center. At columns above $\rm 10^{22}~cm^{-2}$ trapping of Lyman alpha photons becomes effective and consequently the emissivity drops sharply above these columns. Our results are consistent with previous studies \citep{2006ApJ...652..902S,2010ApJ...712L..69S,2011MNRAS.411.1659L}. We also computed the flux emanating from the halo. We find that the flux is of the order of $\rm 10^{-17} - 5 \times 10^{-15}~ erg/cm^{2}/s$ for different background UV fluxes. Similar to the emissivity, the flux is extended and comes from the relatively hot gas in the outskirts of a halo. The Lyman alpha wavelength observed at redshift 5.5 will be 0.79 $\rm \mu m$. This should be detectable with the upcoming telescope JWST. NIRcam/NIRSpec instruments available on JWST will be well suited for its detections. The mass of our halo is $\rm 10^{10}~M_{\odot}$ and has an angular size of 0.52 arc-sec at redshift 5.5, attainable by JWST. Subaru telescope will also be able to detect this flux using suprime-cam in combination with intermediate-band filters (IA679, IA505) as discussed in the previous studies, see \cite{2006ApJ...648...54S}. Our results are in the range of planned surveys in future with Subaru which will make use of next-generation wide-field camera, Hyper Suprime-Cam, and several narrowband filters. We also computed the surface brightness profiles for different background UV fluxes which are shown in figure \ref{sbfig}. The surface brightness profiles are flat as most of the flux emerges from the envelope of a halo rather than its core. They are comparable to the previously observed values \citep{2006ApJ...648...54S,2009ApJ...696.1164O}.

In the presence of large velocity gradients, line trapping of Lyman alpha photons does not remain effective. For $\rm \tau >10^{7}$, only photons far-away in the line wings will be able to escape. Trapping in our case occurs at radii $\rm >10^{20.5}$ cm, where large velocity gradients are significantly reduced. Scattering of Lyman alpha photons which perform a random walk in frequency and space space will blue shift the frequencies leading to an upper limit on the trapping time $\rm \sim t_{ff}(\Delta v/v)$ \citep{1977MNRAS.179..541R}. Although, there are some uncertainties in the diffusion time scale. The issue of trapping has been studied in detail by \cite{2006ApJ...652..902S}. They found that weak dependence of photon escape time on number density ($\rm t_{esc} \propto ~n^{-1/9}$) as compared to the free fall time results in $\rm t_{esc}/t_{ff} \sim n^{7/18}$. Consequently, trapping will remain effective.

We found that the critical value of the background UV flux for $\rm H_{2}$ formation is $\rm J_{21}=1000$, which is in accordance with previous estimates \citep{2001ApJ...546..635O,2003ApJ...596...34B,2007ApJ...671.1559W,2010MNRAS.402.1249S}. We also noticed that the HD abundance remains lower than the critical value ($\rm i.e., 10^{-6}$) for $\rm J_{21} > 0.1$. It could be due to little amount of $\rm H_{2}$ formation and direct photodissociation of HD. Consequently, HD cooling becomes important for lower values of background flux ($\rm i.e., J_{21} < 0.1$). Our results are consistent with the critical value of UV flux for HD derived by \cite{2011MNRAS.412.2603W}. Collisions produce trace amounts of HD at high densities which is too small to dominate the cooling for higher values of $\rm J_{21}$. It has been found that HD cooling is mainly controlled by $\rm H_{2}$ formation rather than direct dissociation. For further details on the role of UV radiation in the formation of HD and its implications for the structure formation see dedicated studies \cite{2008ApJ...685....8M,2011MNRAS.412.2603W}. We also performed simulations to take into account the local variations in the background flux, i.e., $\rm J_{21}=100~and~1000$. We found that for these cases, emission of Lyman alpha photons is reduced, even for an escape fraction of even 1\%, compared to cases with lower values of $\rm J_{21}=0.1 ~and~ 0.01$. However, these are rare occurences, which will only apply in the near vicinity of sources (immediate neighbours). We also studied a few cases for different escape fractions up to 50\%, and found that a higher escape fraction of ionizing photons enhances the emissivity of Lyman alpha photons. We also computed the emission in the helium Lyman alpha line and found that it  is considerably weaker than HI Ly $\alpha$, due to the low abundance of ionized helium in the halo. This may, however, be different in the presence of an AGN or HII regions from massive stars. Further discussion on helium lines can be found in \cite{2006ApJ...640..539Y} and \cite{2000ApJ...537L...5H}.

In our case, ionization predominantly occurs at low densities. Therefore, the contribution of the Lyman alpha recombination emission is a few orders of magnitude lower than the collisionally excited line emission. At radii $\rm 10^{20.5}~cm$, emission diminishes as Lyman alpha line trapping becomes effective in the center. In the case of radiation sources at high densities, recombination emission could be more important. We have further assumed that the background UV flux is constant in time, although we do not expect significant differences if a more detailed model of the time evolution of the UV flux is adopted. \cite{2004ApJ...601..666D} found that the impact of a radiation background is less at high redshift ($\rm z>10$) as compared to $\rm z=3$ due to different properties of the halo.

In light of these results, it will be important to obtain a better understanding of the environmental conditions, in particular the strength of the UV background radiation field, for $\rm \sim 10^{10}~M_{\odot}$ halos close to the reionization epoch. We are further planning to assess additional uncertainties, in particular those related to the metallicity of the gas.

We have assumed here that the halo is metal free for the reasons mentioned in section 3. Indeed pristine halos may exist down to redshift 6 \citep{2009ApJ...700.1672T}. The presence of small amounts of dust may absorb the Lyman alpha photons and re-emit them as continuum at lower frequencies. This can change the scenario. Nevertheless, if the medium is inhomogeneous, escape of Lyman $\alpha$ becomes possible again \citep{1999ApJ...518..138H,1991ApJ...370L..85N}, also see \cite{1996ApJ...468..462G,2006MNRAS.367..979H,2011arXiv1102.1509S}. $\rm H_{2}$ formation is significantly enhanced even for low dust abundances of $\rm 10^{-5}-10^{-4}$ solar \citep{2009A&A...496..365C}. The impact of dust on Lyman alpha emission is an important issue and needs a dedicated study. In an upcoming paper we will explore the influence of dust on the  emission of Lyman alpha in detail. The ubiquity of outflows may even enhance the emission of Lyman alpha photons up to 5\%, by directly injecting them along the line of sight \citep{2010MNRAS.408..352D} while the presence of turbulence may amplify magnetic fields in these protogalaxies, which may affect subsequent star formation \citep{2010A&A...522A.115S}. .

We adopted a background UV flux with a $\rm T_{*}=10^{4}$ K thermal spectrum, typical for radiation dominated by Pop II stars when $\rm z<6$. The $\rm H_{2}$ abundance is mainly controlled by $\rm H^{-}$ photodissociation for a soft UV spectrum in our simulations. For higher radiation temperatures, it will be governed by $\rm H_{2}$ photodissociation. This may have a strong impact on the critical $\rm J_{21}$ values of $\rm H_{2}$ and HD. If Pop III stars with $\rm T_{*}=10^{5}$ K dominate the UV background, then the critical value of $\rm J_{21}$ required for the photo-dissociation of $\rm H_{2}$ increases by two orders of magnitude as computed by previous studies \citep{2003ApJ...596...34B,2007ApJ...671.1559W,2010MNRAS.402.1249S}. On the other hand, for a $\rm T_{*}=10^{5}$ K blackbody spectrum, an almost 100 times smaller value for the ionizing UV background suffices to photo-ionize the gas. So, our estimates may vary for a harder spectrum (i.e., $\rm 10^{5}~K$) but we do not expect a big change in the overall conclusions.

Our results include local radiative transfer effects but nonetheless are comparable to complete radiative transfer studies \citep{2006ApJ...649...14D,2006ApJ...649...37D,2008MNRAS.384.1080T,2008ApJ...685...40W,2008A&A...491...89V,2009MNRAS.393..171M,2010A&A...522A..24H,2010ApJ...725..633F,2010ApJ...716..574Z,2011ApJ...726...38Z}. Lyman alpha radiative transfer strongly depends on the density and velocity profile of a halo due to the resonance nature of the Lyman alpha line and its large scattering cross section. Therefore, we expect some variation in the emergent Lyman alpha flux due to the presence of anisotropies in HI columns along different lines of sight.

We have also ignored the impact of X-rays on the emission of Lyman alpha photons. The presence of X-rays will heat the gas and increase the emissivity of Lyman alpha. Lyman alpha blobs powered by X-ray sources have already been detected \citep[roughly 17\% of the observed sources,][]{2009ApJ...700....1G}. Due to complex chemistry involved in the modeling, X-ray sources demand complete radiative transfer simulations to probe their impact on Lyman alpha emission. We intend to pursue this in the future.


\section*{Acknowledgments}

The FLASH code was in part developed by the DOE-supported Alliance Center for Astrophysical Thermonuclear Flashes (ACS) at the University of Chicago. We thank Robi Banerjee, Christoph Federrath and Seyit Hocuk for discussions on the FLASH code. DRGS acknowledges funding via the European Community Seventh Framework Programme (FP7/2007-2013) under grant agreement No. 229517. Our simulations were carried out on the Gemini machines at the Kapteyn Astronomical Institute, University of Groningen. We thank the anonymous referee for a careful reading of the manuscript and valuable feedback.

\label{lastpage}

\bibliography{biblio4.bib}

\newcommand{\HId}{$\mathrm{H}$}
\newcommand{\HIId}{$\mathrm{H}^+$}
\newcommand{\HzIId}{$\mathrm{H}_2^+$}
\newcommand{\HzId}{$\mathrm{H}_2$}
\newcommand{\HMd}{$\mathrm{H}^-$}
\newcommand{\HeMd}{$\mathrm{He}^-$}
\newcommand{\HeHIId}{$\mathrm{HeH}^+$}
\newcommand{\HeId}{$\mathrm{He}$}
\newcommand{\HeIId}{$\mathrm{He}^+$}
\newcommand{\HeIIId}{$\mathrm{He}^{++}$}
\newcommand{\DId}{$\mathrm{D}$}
\newcommand{\DIId}{$\mathrm{D}^+$}
\newcommand{\HDIId}{$\mathrm{HD}^+$}
\newcommand{\HDId}{$\mathrm{HD}$}
\newcommand{\DMd}{$\mathrm{D}^-$}
\newcommand{\ed}{$\mathrm{e}^-$}



\newpage

\newpage

\appendix

\section{}

\subsection{Chemical Model}

We self-consistently solve the chemical rate equations for our cosmological simulations in an expanding universe. Rate equations are solved for the  following species: $\rm H,~H^{+},~He,~He^{+},~He^{++},~e^{-},~H^{-},~H_{2},~H_{2}^{+},~D,~D^{+},~HD$  under non-equilibrium ionization/equilibrium conditions. We also treat electronic states of hydrogen, up to 5 levels, as separate species. The rate equations are 
\begin{equation}
\rm {dn_{i}\over {dt}} =-D_{i}n_{i}+C_{i} ,
\label{rate1}
\end{equation}
where $\rm n_{i}$ is the number density of the ith specie, and $\rm D_{i}$ and $\rm C_{i}$ are destruction and creation coefficients for the ith specie. Equation \ref{rate1} is solved using the backward differencing scheme (BDF). The solution for equation \ref{rate1} is 
\begin{equation}
\rm n_{i} ={C^{new}dt+ n_{i}^{old} \over{1+ D^{new}dt}} .
\label{rate2}
\end{equation}
Values for $\rm C_{i}^{new}$ and $\rm D_{i}^{new}$ are approximated using species densities from the previous time step and those from the new time step. During each hydrodynamical time step, the chemical time step is computed as
\begin{equation}
\rm dt_{c} = min(HI/(dHI/dt),H^{+}/(dH^{+}/dt))/10 .
\end{equation}
For a given chemical time step, species are evolved until the chemical time step becomes equal to the hydrodynamical time step. We take the initial abundances of species from \cite{2008A&A...490..521S}. These are consistent with nucleosynthesis and observational constraints.

Strong radiation fluxes can photo-ionize the gas, and for a primordial composition specifically hydrogen and helium. Photo-ionization rates are calculated by solving the following equation \citep{1997NewA....2..209A}:
\begin{equation}
\rm {{\partial \rho} \over {\partial t}} / \rho_{k} = \int_{\nu_{th}}^{\infty} 4\pi \sigma_{\nu}^{k} {{I(\nu)} \over {h \nu}} d\nu
\label{ionize}
\end{equation}
where k denotes $\rm H,~He,~He^{+}$, $\rm I(\nu)$ is the intensity of the radiation field, $\rm \nu_{th}$ is the threshold energy for each rate for which ionization takes place and $\rm \sigma_{\nu}^{k}$ is the frequency dependent photo-ionization rate. In order to compute the photo-heating produced due to ionization of the gas the following equation is solved:
\begin{equation}
\rm \Gamma = \int_{\nu_{th}}^{\infty} 4\pi \sigma_{\nu}^{k} {{I(\nu)} \over {h \nu}} (h\nu -h\nu_{th}) d\nu .
\label{ionizeh}
\end{equation}
Details about $\rm H_{2}$ and HD chemistry are given in the section below.

\subsection{$\rm H_{2}$ and HD chemistry}

Neutral hydrogen can capture free electrons, which leads to $\rm H_{2}$ through the $\rm H^{-}$ route. Another channel for $\rm H_{2}$ formation is the collision of H atoms with $\rm H_{2}^{+}$. Similarly, HD can be formed through collisions of $\rm H_{2}$ molecules with deuterium species. We have included collisional and radiative formation and destruction of $\rm H_{2}$ and HD in our network. The relevant rates are listed in table 1. We have also incorporated the photodissociation rates of $\rm H^{-}$, $\rm H_{2}$ and $\rm H_{2}^{+}$ by the background radiation field. Rate equations for $\rm H_{2}$ and HD species are solved under non-equilibrium conditions, while for the ions $\rm H^{-}$ and $\rm H_{2}^{+}$ chemical equilibrium is established rapidly. The reactions involving these species occur on short time scale. The photodissociation and photo-ionization rates of these species are calculated by solving  equation \ref{ionize}, where k denotes $\rm H,~H^{-},~H_{2}^{+}~and~H_{2}$, $\nu_{th}$ is the threshold energy for photo-dissociation  and $\rm \sigma_{\nu}^{k}$ is the frequency dependent photodissociation rate. We also include heating due to photodissociation (for both the Solomon process and direct dissociation) of $\rm H_{2}$ by absorption of back ground radiation in the Lyman-Werner band. Similarly, we compute the photo-dissociation of HD molecules in the presence of a UV field. We take into account the self-shielding of $\rm H_{2}$ in the Lyman-Werner band when computing the $\rm H_{2}$ photodissociation rate. We use the expression given in equation \ref{rate4} \citep{2010MNRAS.402.1249S}. We also include the photodissociation of HD, which is approximately 10\% higher than the $\rm H_{2}$ photodissociation rate for the Solomon process. For direct dissociation they are roughly the same. To compute the $\rm H_{2}$ column density we use the expression from \cite{2010MNRAS.402.1249S} as given in equation \ref{rate5}:

\begin{equation}
\rm f_{sh} = min[1, ({N_{H_{2}} \over {10^{14}~cm^{-2}}})^{-3/4}] ,
\label{rate4}
\end{equation}

\begin{equation}
\rm N_{H_{2}} = f_{H_{2}}n_{tot} \lambda_{J} ,
\label{rate5}
\end{equation}
where $\rm N_{H_{2}}$ is the H$_{2}$ column density, $\rm n_{tot}$ is the total particle density and $\rm \lambda_{J}$ is the Jeans length. $\rm f_{sh}$ is the self-shielding factor to correct the impinging UV flux \citep{1996ApJ...468..269D}. This allows us to compute $\rm N_{H_{2}}$ and $\rm f_{sh}$ from local quantities only.
For molecular hydrogen cooling we have used the cooling function of \cite{1998A&A...335..403G}. We also include heating/cooling rates due to formation and destruction of $\rm H_{2}$ from \cite{2007ApJ...666....1G}. We also include HD photodissociation by a background UV flux in our chemical model. The HD photodissociation rate is adopted from \cite{2007ApJ...666....1G}. For HD cooling we have used a fit to the cooling function of \cite{2005MNRAS.361..850L}.

\subsection{$\rm Ly \alpha$ Trapping}
We have modeled the hydrogen atom as a multi-level system with five levels. Our level population model is based on \cite{2010ApJ...712L..69S}. For the first excited state we distinguish the levels  2s and 2p. Transitions from 2p-1s produce Lyman alpha photons, while transitions from 2s-1s produces the 2-photon continuum. We calculate transition rates from level i to level j using the equation:

\begin{equation}
\rm R_{ij}= A_{ij}\beta_{esc,ij}(1+Q_{ij})+C_{ij}  i>j ,
\end {equation}
\begin{equation}
\rm R_{ij}= {{g_j} \over {g_i}} A_{ji} \beta_{esc,ji}Q_{ji} + C_{ij}, i<j ,
\end{equation}
where $\rm A_{ij}$ are the radiative decay rates, $\rm \beta_{esc,ij}$ the escape probabilities of for transition ij and $\rm C_{ij}$ the collisional excitation and de-excitation rates, adopted from \cite{2001ApJ...546..635O}, $\rm {g_i}$ the statistical weight of the ith level and $\rm Q_{ji}= c^{2} {J_{cont,ij}/ (2h \nu_{ij}^{3})}$, where $\rm \nu_{ij}$ is the frequency of the transition and $\rm J_{cont,ij}$ is the average intensity of the background radiation field. We take the background radiation field to be a black body temperature $\rm 10^{4}$ K \citep{2009MNRAS.393..911G}. The escape probability for the Ly $\rm \alpha$ photons is given by
\begin{equation}
\rm \beta_{esc,ij}= {1-exp(- \tau_{ij}) \over \tau_{ij} } exp({- \beta t_{ph} \over t_{coll}}) ,
\end {equation}
where $\rm \tau_{ij}$ is the optical depth at line center of transition ij, $\rm t_{ph}$ is the photon diffusion time, $\rm t_{coll}$ is the collapse time of the gas, and $\rm \beta$ is a geometrical factor with $\rm \beta$=3 for spherical collapse. Note that $\rm t_{ph} \ll t_{coll}$ for most transitions. For the Lyman alpha line and other direct transitions to the ground states expressions for optical depth and diffusion time are taken from \cite{2006ApJ...652..902S}. For Lyman $\rm \alpha$, $\rm t_{ph}= L(a \tau_{21})^{1/3}/c$, where $a$ is natural-to-thermal line width and the optical depth $\rm \tau_{21}= 1.04 \times 10^{-13} N_{H}T_{4}^{-0.5}$. The cooling function for hydrogen lines is computed from the escape probabilities, level populations and strength of the background radiation field. Further details of all these processes can be found in \cite{2001ApJ...546..635O} and \cite{2010ApJ...712L..69S,1977MNRAS.179..541R,2006ApJ...652..902S,2011MNRAS.411.1659L,2010ApJ...712L..69S}.

\begin{table*}
 \centering
\caption{~~~~~~~~~~~~~~~~~~~~~~~~~~~~~ \textbf{Collisional and radiative rates}}
\begin{tabular}{llll}

\hline\hline
No. & Reaction & Rate coefficient $\rm cm^{3} s^{-1}$/Cross-section  $\rm \sigma (cm^{2})$ & Reference \\ \hline

 1 &  \HId ~+ \ed $\rightarrow$   \HIId ~+ 2\ed  & \textbf{see reference}   &     AAZN97  \\ 
 2  &  \HIId ~+ \ed   $\rightarrow$   \HId ~+ $\rm \gamma$ & \textbf{see reference} & AAZN97  \\ 
 3  &  \HeId ~+ \ed   $\rightarrow$   \HeIId ~+ 2\ed   & \textbf{see reference} & AAZN97  \\ 
 4  &  \HeIId  ~+  \ed   $\rightarrow$   \HeId ~+ $\rm \gamma$   & \textbf{$\rm 3.92\times10^{-13}/T_{e}^{0.6353}$} for \textbf{$\rm T_{e} < 8000$}   & AAZN97  \\ 
 &  &  \textbf{  $\rm 1.54\times10^{-9}T^{-1.5} exp({-44.569~eV \over T_{e}})[0.3 ~+ exp({8.10~eV \over T_{e}})]$} for \textbf {$\rm T_{e} > 8000 $}  & AAZN97 \\
 5  &  \HeIId ~+ \ed $\rightarrow$   \HeIIId ~+ 2\ed  & \textbf{see reference}   &     AAZN97  \\ 
 6  &  \HeIIId ~+ \ed $\rightarrow$   \HeIId ~+  $\rm \gamma$  & \textbf{$\rm 3.36 \times 10^{-10} T^{-0.5}({T \over 1000})^{-0.2} (1 ~+ ({T \over 10^{6})^{0.7}})^{-1} $} & AAZN97 \\ 
 7  &  \HId  ~+  \ed $\rightarrow$ \HMd ~+ $\rm \gamma$ & \textbf{$\rm dex[-17.845 ~+ 0.762(\log T) ~+ 0.1523(\log T)^{2} -0.03274(\log T)^{3}]$} for \textbf{$\rm T \leq 6000 $} &  GJ07 \\
 &  & \textbf{ $\rm dex[-16.42 ~+ 0.1998(\log T)^{2} -5.447 \times 10^{-3}(\log T)^{4} ~+ 4.0415 $}   \\
 &  & \textbf{$\times 10^{-5}(\log T)^{6} ]$} for \textbf{ $\rm T > 6000$ } &  GJ07 \\
 8  &  \HMd ~+ \HId $\rightarrow$ \HzId ~+ e$^-$ & \textbf{$\rm 1.5\times 10^{-9} $} for \textbf{$\rm T \leq 300 $} &  GJ07 \\
 &  &  \textbf{$\rm 4.0\times 10^{-9} T^{-0.17} $} for \textbf{$\rm T \leq 300 $} for \textbf{$\rm T > 300 $} &  GJ07 \\
 9  &  \HId  ~+  \HIId $\rightarrow$ \HzIId ~+ $\rm \gamma$ & \textbf{$\rm dex[-19.38 -1.523(\log T) ~+ 1.118(\log T)^{2} -0.1269(\log T)^{3}]$}  &  GJ07 \\
 10  &  \HId ~+ \HzIId $\rightarrow$ \HzId ~+ \HIId & \textbf{$\rm 6.4\times 10^{-10} $}  &  GJ07 \\
 11  &  \HMd ~+ \HIId $\rightarrow$ \HId ~+ \HId & \textbf{$\rm 5.4 \times 10^{-6} T^{-0.5} +6.3 \times 10^{-8} -9.2 \times 10^{-11} T^{0.5} +4.4 \times 10^{-13} T $}  &  GJ07 \\
 12  &  \HzIId ~+ \ed $\rightarrow$ \HId ~+ \HId & \textbf{$\rm 1.0 \times 10^{-8} $} for \textbf{ $\rm T \leq 617$ } &  GJ07 \\
 &  &   \textbf{$\rm 1.32 \times 10^{-6} T^{-0.76}$} for \textbf{ $\rm T > 617$ } &  GJ07 \\
 13  &  \HzId ~+ \HIId $\rightarrow$ \HzIId ~+ \HId & \textbf{see reference} &  GJ07 \\
 14 &  \HzId ~+ \ed $\rightarrow$ \HId ~+ \HId ~+ \ed & \textbf{$\rm 3.73 \times 10^{-9} T^{0.1121} exp({-99430 \over T})$} &  GJ07 \\
 15 &  \HzId ~+ \HId $\rightarrow$ \HId ~+ \HId ~+ \HId & \textbf{$\rm 6.67 \times 10^{-12} T^{0.5} exp(-1 ~+ {63590 \over T})$} &  GJ07 \\
 16 &  \HzId ~+ \HzId $\rightarrow$ \HzId ~+ \HId ~+ \HId & \textbf{$\rm {5.996 \times 10^{-30} T^{4.1881} \over (1.0 +6.761 \times 10^{-6} T)^{5.6881}}exp(-1 ~+ {63590 \over T})$} & GJ07 \\
 17 &  \DId ~+ \ed $\rightarrow$ \DIId ~+ \ed ~+ \ed & \textbf{see reference} &  GJ07 \\
 18 &  \DIId ~+ \ed $\rightarrow$ \DId ~+ $\rm \gamma$ & \textbf{$\rm 2.753 \times 10^{-14}  ({315614 \over T})^{1.5}   ((1.0+ {115188 \over T})^{0.407})^{-2.242}$} &  GJ07 \\
 19 &  \HMd ~+ \ed $\rightarrow$ \HId ~+ \ed ~+ \ed & \textbf{see reference} &  GJ07 \\
 20 & \HMd ~+ \HId $\rightarrow$ \HId ~+ \HId ~+ \ed & \textbf{see reference} &  GJ07 \\
 21 & \HMd ~+ \HId $\rightarrow$ \HzId ~+ \ed & \textbf{$\rm 6.9 \times 10^{-9} T^{-0.35}$} for \textbf{$\rm T \leq 8000 $} &  GJ07 \\
 &  & \textbf{$\rm 9.6 \times 10^{-7} T^{-0.90}$} for \textbf{$\rm T > 8000 $} &  GJ07 \\
 22 &  \HId ~+ \DIId $\rightarrow$ \DId ~+ \HIId & \textbf{$\rm 2.06 \times 10^{-10} T^{0.396} exp({-33 \over T}) ~+ 2.03 \times 10^{-9} T^{-0.332} $} &  GJ07 \\
 23 &  \DId ~+ \HIId $\rightarrow$ \HId ~+ \DIId & \textbf{$\rm 2.0 \times 10^{-10} T^{0.402} exp({-37.1 \over T}) -3.31 \times 10^{-17} T^{1.448} $} for \textbf{$\rm T \leq 2.0 \times 10^{5} $} &  GJ07 \\
  & &  \textbf{$\rm 3.44 \times 10^{-10} T^{0.35}$} for \textbf{$\rm T > 2.0 \times 10^{5} $} &  GJ07 \\
 24 &  \HzId ~+ \DIId $\rightarrow$ \HDId ~+ \HIId & \textbf{$\rm [0.417 +0.864 \log T -0.137 (\log T)^{2}] \times 10^{-9} $} &  GJ07 \\
 25 &  \HDId ~+ \HIId $\rightarrow$ \HzId ~+ \DIId & \textbf{$\rm 1.1 \times 10^{-9} exp({-488 \over T}) $} &  GJ07 \\
 26 &  \HzId ~+ \DId $\rightarrow$ \HDId ~+ \HId & \textbf{$\rm 1.69 \times 10^{-10} exp({-4680 \over T}) $} for \textbf{ $\rm T \leq 200$ } &  GJ07 \\
 &  &   \textbf{$\rm 1.69 \times 10^{-10} exp({-4680 \over T} ~+ {198800 \over T^{2}}) $} for \textbf{ $\rm T > 200$ } &  GJ07 \\
 27  &  \HDId ~+ \HId $\rightarrow$ \DId ~+ \HzId & \textbf{$\rm 5.25 \times 10^{-11} exp({-4430 \over T}) $} for \textbf{ $\rm T \leq 200$ } &  GJ07 \\
 &  &  \textbf{$\rm 5.25 \times 10^{-11} exp({-4430 \over T} ~+ {173900 \over T^{2}}) $} for \textbf{ $\rm T > 200$ }&  GJ07 \\
 28 & \HMd ~+ $\rm \gamma$ $\rightarrow$ \HId ~+ \ed & \textbf{$\rm \sigma_{23}=7.928 \times 10^{5}(\nu -\nu_{th})^{3 \over 2} ({1 \over \nu })^{1 \over 3} cm^{2}$}  for \textbf{$\rm h\nu > h\nu_{th} =0.775~eV$} & AAZN97 \\
29 & \HzId ~+ $\rm \gamma$ $\rightarrow$ \HzIId ~+ \ed & \textbf{$\rm \sigma_{24}= 0 $}  for \textbf{$\rm h\nu < 15.42~eV$}  & AAZN97 \\
& &  \textbf{$\rm \sigma_{24}= 6.2 \times 10^{-18}h\nu -9.4 \times 10^{-17} cm^{2}$}  for \textbf{$\rm 15.42 \leq h\nu < 16.50~eV$} & AAZN97 \\
& &  \textbf{$\rm \sigma_{24}= 1.4 \times 10^{-18}h\nu -1.48 \times 10^{-17}cm^{2}$}  for \textbf{$\rm 16.50 \leq h\nu < 17.7~eV$} & AAZN97 \\
& &  \textbf{$\rm \sigma_{24}= 0 $}  for \textbf{$\rm h\nu \geq 17.7~eV cm^{2}$} for \textbf{$\rm  h\nu \geq 17.7~eV$} & AAZN97 \\
30 & \HzIId ~+ $\rm \gamma$ $\rightarrow$ \HId ~+ \HIId & \textbf{$\rm \sigma_{25}= dex[-40.97 ~+ 15.9795({E \over T_{th}}) -3.53934({E \over T_{th}})^{2} +0.2581155({E \over T_{th}})^{3}] $}   \\
&  &   for \textbf{$\rm E_{th} =2.65~ eV 2.65 < E < 11.27~eV$} &  GJ07 \\
&  &  \textbf{ $\rm dex[-30.26 ~+ 7.3935({E \over T_{th}}) -1.2914({E \over T_{th}})^{2} ~+ 6.5785 \times 10^{-2}({E \over T_{th}})^{3}] $} \\
& & for \textbf{$\rm 11.27 < E < 21.0~eV$} & GJ07  \\
 31 & \HzIId ~+ $\rm \gamma$ $\rightarrow$ 2$\rm \times$\HId ~+ \ed   & AAZN97 \\
& & \textbf{$\rm \log(\sigma_{26})={-16.926-4.528 \times 10^{2} h\nu ~+ 2.238 \times 10^{-4}(h \nu)^{2} ~+ 4.245 \times 10^{-7}(h \nu)^{3}}$} \\
& & for \textbf{$\rm 30~eV <h\nu < 90~eV$}  & AAZN97 \\
32 & \HzId ~+ $\rm \gamma$ $\rightarrow$ \HzId* $\rightarrow$ \HId ~+ \HId & \textbf{$\rm 1.12 \times 10^{8} J(\nu)$} where J the is flux at E=12.87 eV & AAZN97 \\
33 & \HzId ~+ $\rm \gamma$ $\rightarrow$ \HId ~+ \HId & \textbf{$\rm \sigma_{28}= {1 \over y+1}(\sigma_{28}^{L0} ~+ \sigma_{28}^{W0}) 1.0- {1 \over y+1}(\sigma_{28}^{L1} ~+ \sigma_{28}^{W1})$} for sigma values see reference  & AAZN97 \\
34 & \HId ~+ $\rm \gamma$ $\rightarrow$ \HIId ~+ \ed & \textbf{$\rm \sigma_{20}= {A_{0} \over Z^{2}} ({\nu \over \nu_{th}})^{4} 
    {{exp[4.0- 4.0(arctan\epsilon)/ \epsilon]}   \over {1-exp(-2\pi/ \epsilon)}} $}  & AAZN97 \\
35 & \HeIId ~+ $\rm \gamma$ $\rightarrow$ \HeIIId ~+ \ed & \textbf{$\rm \sigma_{22}= {A_{0} \over Z^{2}} ({\nu \over \nu_{th}})^{4} 
    {{exp[4.0- 4.0(arctan\epsilon)/ \epsilon]}   \over {1-exp(-2\pi/ \epsilon)}} $}  & AAZN97 \\
36 & \HeId ~+ $\rm \gamma$ $\rightarrow$ \HeIId ~+ \ed & \textbf{$\rm \sigma_{21}=  7.42 \times 10^{-18}(1.66({\nu \over \nu_{th}})^{-2.05} 
    -0.66({\nu \over \nu_{th}})^{-3.05} $}  & AAZN97 \\
37 & \HDId ~+ $\rm \gamma$ $\rightarrow$ \HId ~+ \DId & \textbf{see reference}   & GJ07 \\
\hline 
\end{tabular}
\label{Table1}

\flushleft 
In the table above, $\rm T$ denotes the gas temperature in Kelvin, $\rm T_{e}$ the gas temperature in eV, $\rm T_r$ the temperature of radiation in K. $\mathrm{dex}(x)=10^x$. Acronyms: AAZN97: Abel, Anninos, Zhang, Norman (1997); GJ07: Glover and Jappsen (2007).
\end{table*}


\begin{table*}

\caption{~~~~~~~~~~~~~~~\textbf{Heating/cooling rates} }

\begin{tabular}{llll}

 \hline\hline
Process  & Rate  & Reference   \\ \hline
&      Heating/Cooling & \\ \hline

Collisional excitation cooling  &  \textbf{$\rm 7.5 \times 10^{-19}(1 ~+ \sqrt{T_{5}})^{-1} exp(-11838/T)n_{e}n_{H}$}   &     AZAN96  \\ 
&  \textbf{$\rm 9.1 \times 10^{-27}(1 ~+ \sqrt{T_{5}})^{-1} T^{-0.1687}exp(-11838/T)n_{e}^{2}n_{He}$}   &       \\ 
& \textbf{$\rm 5.54 \times 10^{-17}(1 ~+ \sqrt{T_{5}})^{-1} T^{-0.397}exp(-473638/T)n_{e}n_{He}^{+}$}   &       \\ 

Collisional ionization cooling  &  \textbf{$\rm 2.18 \times 10^{-11}k_{1}n_{e}n_{H}$}   &     AZAN96  \\ 
&  \textbf{$\rm 2.18 \times 10^{-11}k_{1}n_{e}n_{H}$}   &      \\ 
&  \textbf{$\rm 3.94 \times 10^{-11}k_{3}n_{e}n_{He}$}   &      \\
&  \textbf{$\rm 8.72 \times 10^{-11}k_{5}n_{e}n_{He}^{+}$}   &       \\
&  \textbf{$\rm 5.01 \times 10^{-27}(1 ~+ \sqrt{T_{5}})^{-1}T^{-0.1687}exp(-55338/T)n_{e}^{2}n_{He}^{+}$}   &      \\

Recombination cooling  &  \textbf{$\rm 8.7 \times 10^{-27}{T}^{1/2}T_{3}^{-0.2}(1 ~+ T_{6}^{0.7})^{-1}n_{e}n_{H}^{+}$}   &     AZAN96  \\ 
&  \textbf{$\rm 1.55 \times 10^{-26}T^{0.3647}n_{e}n_{He}^{+}$}   &    \\ 
&  \textbf{$\rm 1.24 \times 10^{-13}T^{-1.5}[1.0+ 0.3exp(-94000/T)]exp(-47000/T)n_{e}n_{He}^{+}$}   &    \\ 
&  \textbf{$\rm 3.48 \times 10^{-26}T^{1/2}T_{3}^{-0.2}(1.0 +T_{6}^{0.7})^{-1}n_{e}n_{He}^{++}$}   &    \\

Bremsstrahlung cooling  & \textbf{$\rm 1.43 \times 10^{-27}\sqrt{T}[1.1 ~+ 0.34 exp(-(5.5- \log(T))^{2}/3)]n_{e}(n_{H}^{+} ~+ n_{He}^{+} ~+ n_{He}^{++})$}   &    AZAN96  \\ 

Compton cooling  & \textbf{$\rm 5.65 \times 10^{-36}(1+z)^{4}[T - 2.75(1+z)]n_{e}$}   &    AZAN96  \\ 

Molecular hydrogen cooling  & \textbf{see reference}   &    GP98  \\ 

HD cooling  & \textbf{see reference}  & LNA05 \\ 
$\rm H_{2}$ Photodissociation heating  &  & AAZ97 \\
 Solomon process  &  \textbf{$\rm \Gamma = 6.4 \times 10^{-13} H_{2}k27$}   \\

 Direct heating  &  \textbf{$\rm \Gamma = H_{2}\int_{\nu_{th}}^{\infty}4 \pi \sigma_{28}(\nu){ i(\nu) \over h \nu }(h \nu -h \nu_{th} )d \nu$}   \\
Gas-phase $\rm H_{2}$ formation heating & \textbf{$\rm \Gamma = 2.93 \times 10^{-12}k_{2} n_{H}^{-} ~+ 5.65 \times 10^{-12} k_{4} n_{H_{2}^{+}} n_{H}({n \over (n ~+ n_{cr})})$} & GJ07 \\

$\rm H_{2}$ Collisional dissociation cooling & \textbf{$\rm \Lambda = 7.2 \times 10^{-12} (k_{9}n_{H} ~+ k_{10}n_{H_{2}})n_{H_{2}}$} & GJ07  \\

Cooling from electronic states of hydrogen & \textbf{see reference} & SSG10  \\
\hline 

\end{tabular}

\label{Table2}

\flushleft 
In the table above, $\rm T$ denotes the gas temperature in Kelvin, $\rm T_{n}$ the gas temperature in $\rm 10^{n}$, $\rm T_r$ the temperature of radiation in K. $\mathrm{dex}(x)=10^x$. Acronyms: AAZN97: Abel, Anninos, Zhang, Norman (1997); GJ07: Glover and Jappsen (2007); GP98: Galli and Palla (1998); AZAN96: Anninos, Zhang, Abel, Norman (1996); LNA05: Lipovka, A. et al. (2005); SSG10: Schleicher, Spaans, Glover (2010).

\end{table*}

\end{document}